\documentclass[twocolumn,bbm, twocolappendix, trackchanges]{aastex631} 

\usepackage{graphicx}
\usepackage{amsmath,amssymb}
\usepackage[section]{placeins}
\usepackage{hyperref}

\usepackage{soul}
\usepackage{threeparttablex}
\usepackage{multirow}
\usepackage{stfloats}
\usepackage{amsmath}
\usepackage{tablefootnote} 
\usepackage{xcolor}
\usepackage{float}

\DeclareOldFontCommand{\bf}{\normalfont\bfseries}{\mathbf} 
\providecommand{\DIFdeltex}[1]{} 
\providecommand{\DIFaddbegin}{} 
\providecommand{\DIFaddend}{} 
\providecommand{\DIFdelbegin}{} 
\providecommand{\DIFdelend}{} 
\providecommand{\DIFaddbeginFL}{} 
\providecommand{\DIFaddendFL}{} 
\providecommand{\DIFdelbeginFL}{} 
\providecommand{\DIFdelendFL}{} 
\newcommand{\DIFscaledelfig}{0.5}
\RequirePackage{settobox} 
\RequirePackage{letltxmacro} 
\newsavebox{\DIFdelgraphicsbox} 
\newlength{\DIFdelgraphicswidth} 
\newlength{\DIFdelgraphicsheight} 
\LetLtxMacro{\DIFOincludegraphics}{\includegraphics} 
\newcommand{\DIFaddincludegraphics}[2][]{{\color{blue}\fbox{\DIFOincludegraphics[#1]{#2}}}} 
\newcommand{\DIFdelincludegraphics}[2][]{
\sbox{\DIFdelgraphicsbox}{\DIFOincludegraphics[#1]{#2}}
\settoboxwidth{\DIFdelgraphicswidth}{\DIFdelgraphicsbox} 
\settoboxtotalheight{\DIFdelgraphicsheight}{\DIFdelgraphicsbox} 
\scalebox{\DIFscaledelfig}{
\parbox[b]{\DIFdelgraphicswidth}{\usebox{\DIFdelgraphicsbox}\\[-\baselineskip] \rule{\DIFdelgraphicswidth}{0em}}\llap{\resizebox{\DIFdelgraphicswidth}{\DIFdelgraphicsheight}{
\setlength{\unitlength}{\DIFdelgraphicswidth}
\begin{picture}(1,1)
\thicklines\linethickness{2pt} 
{\color[rgb]{1,0,0}\put(0,0){\framebox(1,1){}}}
{\color[rgb]{1,0,0}\put(0,0){\line( 1,1){1}}}
{\color[rgb]{1,0,0}\put(0,1){\line(1,-1){1}}}
\end{picture}
}\hspace*{3pt}}} 
} 
\LetLtxMacro{\DIFOaddbegin}{\DIFaddbegin} 
\LetLtxMacro{\DIFOaddend}{\DIFaddend} 
\LetLtxMacro{\DIFOdelbegin}{\DIFdelbegin} 
\LetLtxMacro{\DIFOdelend}{\DIFdelend} 
\DeclareRobustCommand{\DIFaddbegin}{\DIFOaddbegin \let\includegraphics\DIFaddincludegraphics} 
\DeclareRobustCommand{\DIFaddend}{\DIFOaddend \let\includegraphics\DIFOincludegraphics} 
\DeclareRobustCommand{\DIFdelbegin}{\DIFOdelbegin \let\includegraphics\DIFdelincludegraphics} 
\DeclareRobustCommand{\DIFdelend}{\DIFOaddend \let\includegraphics\DIFOincludegraphics} 
\LetLtxMacro{\DIFOaddbeginFL}{\DIFaddbeginFL} 
\LetLtxMacro{\DIFOaddendFL}{\DIFaddendFL} 
\LetLtxMacro{\DIFOdelbeginFL}{\DIFdelbeginFL} 
\LetLtxMacro{\DIFOdelendFL}{\DIFdelendFL} 
\DeclareRobustCommand{\DIFaddbeginFL}{\DIFOaddbeginFL \let\includegraphics\DIFaddincludegraphics} 
\DeclareRobustCommand{\DIFaddendFL}{\DIFOaddendFL \let\includegraphics\DIFOincludegraphics} 
\DeclareRobustCommand{\DIFdelbeginFL}{\DIFOdelbeginFL \let\includegraphics\DIFdelincludegraphics} 
\DeclareRobustCommand{\DIFdelendFL}{\DIFOaddendFL \let\includegraphics\DIFOincludegraphics} 

\usepackage{placeins}
\begin{document}

\title[]
{Co-evolution of bar and spiral arms in TNG50 simulations using Information Theory}

\author{Anagha A G}
\affiliation{Department of Physics, Indian Institute of Science Education and Research (IISER) Tirupati, Tirupati - 517619, India}
\email{anaghaag@students.iisertirupati.ac.in}

\author{Arunima Banerjee}
\affiliation{Department of Physics, Indian Institute of Science Education and Research (IISER) Tirupati, Tirupati - 517619, India}
\email{arunima.iisert@gmail.com}

\begin{abstract}
Using Information Theory, we investigate the co-evolution of bars and spiral arms in barred-spiral galaxies from the cosmological magneto-hydrodynamic Illustris TNG50 simulations. We first calculate Mutual Information (MI) between a structural or kinematic parameter of the bar (bar strength $A_{2bar}$, bar length $r_{bar}$, bar pattern speed $\Omega$) and that of a spiral arm (spiral strength $A_{2spiral}$, spiral arm pitch angle $\Psi$). We calculate MI in two different galaxy samples: (i) one forming bars before spirals (ii) other forming spirals before bars. We note, spirals form immediately after bars in the first sample, whereas bars form 1.7 Gyrs after spirals in the second. We find a high mean MI value in each of these samples  (0.4 - 0.5), and in the combined sample (0.4-0.8), confirming a fair degree of association of the bar and the spiral arm. To identify whether the bar or the spiral arm effectively drives their co-evolution, we calculate the Transfer Entropy (TE) (bar-to-spiral TE, spiral-to-bar TE), from the time series data of each of the above bar-spiral parameter pairs. We find that the median bar-to-spiral TE and spiral-to-bar TE values vary between $ 0.33$ and $ 0.42$ for each galaxy sample, comparable to those of the combined sample. A similar trend was observed in our calculated Liang information flow rates. Our novel approach may possibly indicate that the bar and the spiral arm regulate their co-evolution on an equal footing.

\end{abstract}




\section{Introduction} \label{intro}

More than half of the spiral galaxies are barred-spirals, and our galaxy, the Milky Way itself, is a classic example of a barred-spiral galaxy. These non-axisymmetric structures, bars and spiral arms, play an important role in the secular evolution of galaxies (\citeauthor{burton_1988} \citeyear{burton_1988}, \citeauthor{Benjamin_2005} \citeyear{Benjamin_2005}, \citeauthor{Dame_2001} \citeyear{Dame_2001}). The quadrupole moment
associated with the gravitational potential of the non-
axisymmetric disk components like bars and spiral arms helps in angular momentum transport (\citeauthor{LyndenBellKalnajs1972} \citeyear{LyndenBellKalnajs1972}, \citeauthor{sellwood_binney_2002} \citeyear{sellwood_binney_2002}), and hence in gas inflow, thus fueling star formation activity (\citeauthor{MartinetFriedli1997} \citeyear{MartinetFriedli1997}, \citeauthor{Aguerri1999} \citeyear{Aguerri1999}, \citeauthor{SilvaLima2022} \citeyear{SilvaLima2022}, \citeauthor{Storchi_Bergmann_2019} \citeyear{Storchi_Bergmann_2019}, \citeauthor{kataria_2024} \citeyear{kataria_2024}, \citeauthor{garland2024} \citeyear{garland2024}, \citeauthor{vogel1988} \citeyear{vogel1988}, \citeauthor{kim2002} \citeyear{kim2002}). In addition, being an overdense region in a smooth and homogenous disk, they scatter stars off their orbits and hence lead to heating the stellar disk, which enhances the disk dynamical stability (\citeauthor{Fujii_2011} \citeyear{Fujii_2011}, \citeauthor{Aditya_2025} \citeyear{Aditya_2025}).

According to the Density Wave Theory, the bars and spirals are formed by the density waves and grow by the mechanism of swing amplification (\citeauthor{BinneyTremaine1987} \citeyear{BinneyTremaine1987}). Besides, bars can be associated with spiral arms and may themselves drive the spiral density waves (\citeauthor{BinneyTremaine1987} \citeyear{BinneyTremaine1987}). A strong dynamical coupling between bars and spiral arms is also predicted by the manifold theory, which is an alternative framework for understanding the formation of spirals and rings (\citeauthor{Romero_G_mez_2007} \citeyear{Romero_G_mez_2007}, \citeauthor{patsis_2006} \citeyear{patsis_2006}, \citeauthor{voglis2006} \citeyear{voglis2006}, \citeauthor{Athanassoula2009} \citeyear{Athanassoula2009}). The stellar dynamics in the galactic bar is nonlinear, sustained by mostly chaotic orbits (\citeauthor{SellwoodWilkinson1993} \citeyear{SellwoodWilkinson1993}). The Lagrangian associated with the bar potential plays an important role in the dynamics of these stellar orbits. Non-axisymmetric bar potentials have two unstable points (L1, L2), and two other points (L4, L5) (\citeauthor{ContopoulosandGrosbol1989;} \citeyear{ContopoulosandGrosbol1989;}). The unstable Lagrange points L1 and L2, located at the ends of the bar, are responsible for the formation of spirals, and also strong bars have more unstable Lagrangian points compared to weak bars (\citeauthor{Athanassoula2009} \citeyear{Athanassoula2009}). The chaotic orbits in the corotation region of the strongly-barred galaxies constitute the main building blocks of the spiral arms (\citeauthor{Mondal_2021} \citeyear{Mondal_2021}).

The bar can be modelled as a global, density wave mode with $m=2$, mostly in the region of the disk with a solid-body rotation curve (\citeauthor{SellwoodWilkinson1993} \citeyear{SellwoodWilkinson1993}). A large bar strength corresponds to a large quadrupole moment associated with its gravitational potential, which excites spiral density waves outside the corotation and shapes the orbital families. Also, the bar strength indicates how strongly a bar can trap stars on the elongated $x_{1}$ family of orbits, which supports the bar and induces chaos. Stronger bars exert greater tangential forces, influence the transfer of angular momentum within the disk \citep{Laurikainen_2004, Das_2008} and drive more prominent resonances in the galactic disks (Inner Lindblad ILR, Corotation CR, Outer Lindblad OLR). This, in turn, regulates the angular momentum redistribution among the bar, the disk, and the halo. This angular momentum redistribution drives gas inflows towards the galactic centre, potentially fueling central star formation and active galactic nuclei (AGN) activity \citep{Oh_2012,refId0, Newnham_2020, George_2019, Hogarth_2024, Galloway_2015}. More passive galaxies often host longer bars \citep{Fraser_2019}. The growth of bars, in terms of length and strength, has been studied in numerical simulations. As the host galaxy evolves, the bar radius increases at a pace comparable to that of the disk scale length (\citeauthor{Rosas-Guevara_2022} \citeyear{Rosas-Guevara_2022}). \cite{Romeo_2022} found that galaxies hosting long bars tend to have lower specific angular momentum values compared to those with short bars or unbarred galaxies. It is now well-known that galactic bars rotate about the galactic centre like a solid body with a constant pattern speed $\Omega_{p}$. If $\Omega_{p}$ is too high or too low, the $x_1$ orbits lose their coherence \citep{ContopoulosandGrosbol1989;}. The magnitude of $\Omega_{p}$ is regulated by the dark matter mass fraction in the inner regions of host galaxies, as dark matter can slow down bars through dynamical friction \citep{Weinberg_1985}. This slowing indicates angular momentum transfer via resonances to the halo and outer disk. Interestingly, studies of bars produced in ILLUSTRIS TNG50 indicate that the median value of pattern speed increases with redshift \citep{Semczuk_2024, Habibi_2024}

The spiral arm can also be looked upon as a global stationary wave with $m=2$, but in the sheared disk \citep{Linshu_1964}. The pitch angle ($\alpha$), which quantifies the tightness of the winding of the spiral arm, is defined as the angle between the tangent to a spiral arm and a circle of constant radius from the galactic centre (\citeauthor{Davis_2013} \citeyear{Davis_2013}; \citeauthor{ Michikoshi_2014} \citeyear{ Michikoshi_2014}). A smaller pitch angle indicates more tightly wound spiral arms, while a larger angle indicates more open structures \citep{Davis_2013}. For instance, the pitch angle tends to decrease (i.e., arms are more tightly wound) in galaxies with earlier Hubble types, more prominent bulges, higher stellar concentration, and larger total stellar masses ($M_{\ast}^{\rm gal}$). In the Lin-Shu density wave theory, the pitch angle is expected to remain constant over time, reflecting an inherent property of the underlying galaxy \citep{Pringle_2019}. Conversely, theories involving tidal interactions or self-gravity predict that spiral arms should wind up over time. Both observational data and $N$-body simulations indicate a correlation between pitch angle and galactic shear rate \citep{Grand_2013}, which describes how differential rotation stretches material in the disk. Stronger spiral arms, i.e., spiral features with large enough amplitude of the $m=2$ mode, are more effective at driving gas towards the galactic centre and inducing star formation \citep{Seo_2014, Yu_2021}. Hydrodynamic simulations show that spiral arms can efficiently transport gas from the outskirts to central nuclear rings, provided the arms rotate more slowly than the bar, thereby enhancing star formation in those regions \citep{Seo_2014}. The impact of spiral structure on global star formation has been investigated, showing that arm strength correlates well with variations in star formation rate (SFR) as a function of stellar mass \citep{Yu_2021}. Stronger arms are found in galaxies above the star-forming main sequence, suggesting that spiral structures boost star formation \citep{Yu_2021}.

Numerous studies have examined the correlation between various bar and spiral parameters in galaxies. Theoretical calculations suggest that the central mass concentration and the galactic shear influence the pitch angle of spiral arms. The galactic shear is most closely associated with the bar pattern speed (\citeauthor{Garma_Oehmichen_2021} \citeyear{Garma_Oehmichen_2021}) and bar strength (\citeauthor{schwarz_1984} \citeyear{schwarz_1984}). Interestingly, observational studies indicate that the distribution of pitch angles for barred and unbarred-spiral galaxies is similar (\citeauthor{voglis2006} \citeyear{voglis2006}), and bar-induced gravitational torque and pitch angle are not correlated (\citeauthor{García_2019} \citeyear{García_2019}). Other studies have revealed that the strength of the spiral arms correlates well with the quadrupole moment of the gravitational potential of the bar (\citeauthor{Garma_Oehmichen_2021} \citeyear{Garma_Oehmichen_2021}) and tangential force exerted by the bar(\citeauthor{Salo_2010} \citeyear{Salo_2010}, \citeauthor{García_2019} \citeyear{García_2019}). Finally, though stronger spirals are associated with stronger bars in general, two strongly barred galaxies (NGC 7513 and UGC 10862) exhibit weak spiral amplitudes (\citeauthor{schwarz_1984} \citeyear{schwarz_1984}, \citeauthor{Buta_2009} \citeyear{Buta_2009}). Bars might play a role in creating spirals (\citeauthor{sanders_1976} \citeyear{sanders_1976}), but mainly in strongly barred galaxies (\citeauthor{sellwood_1988} \citeyear{sellwood_1988}). The spiral arms are a result of spiral density waves, which could be driven by bars.

 Despite extensive theoretical and observational studies, several key questions remain unresolved: whether bars and spirals are dynamically coupled, which bar and spiral properties are most strongly linked, whether these relationships are linear or nonlinear, and whether any directional influence exists between their temporal evolutions. These unresolved issues directly motivate the use of information-theoretic methods in the present work.    In this paper, we use Information Theory to study the interdependence between the different pairs of physical parameters of the bar (bar strength $A_{2bar}$, bar length $r_{bar}$, bar pattern speed $\Omega$) and the spiral arm (strength of the spiral arm, $A_{2spiral}$, pitch angle of spiral arm $\Psi$) in the barred-spiral galaxy sample from the TNG50 cosmological magneto-hydrodynamical simulation (\citeauthor{Nelson2019b} \citeyear{Nelson2019b}, \citeauthor{Pillepich2019} \citeyear{Pillepich2019}). Information Theory provides a framework for identifying relationships, including nonlinear dependencies, between variables. Several techniques have been developed using Information Theory to determine the relationships between two random variables. One such technique is \textit{Mutual Information}(MI), which is a model-free method that quantifies the information one random variable can reveal about another (\citeauthor{Thommas_2006} \citeyear{Thommas_2006}). We use MI to study the relation between the physical parameters of the bar and the spiral arm. To assess causal interactions and the direction of the flow of information, we calculate Transfer Entropy (TE). TE is a concept that retains some of the properties of MI and also encapsulates the dynamics of information transfer \citep{ Schreiber_2000}. We quantify the exchange of information between the pair of bar and spiral arm parameters in both directions using TE. Further, we calculate Liang information flow rate (IFR), which also measures the transfer of information between the time series of a pair of bar and spiral parameters (\citeauthor{liang_kleeman_2005} \citeyear{liang_kleeman_2005}, \citeauthor{Liang_2008} \citeyear{Liang_2008}, \citeauthor{liang_2013} \citeyear{liang_2013}, \citeauthor{liang_2014} \citeyear{liang_2014}, \citeauthor{liang_2015} \citeyear{liang_2015}). To our best knowledge, it is the first such attempt in the study of barred spiral galaxies.

 The paper is organised as follows: In \S 2, we introduce the bar and spiral parameters considered in this study, and in \S 3, we present our galaxy sample. In \S 4, we describe Information Theory techniques used for analysis; in \S 5, we present the results and discussion, followed by conclusions in \S 6.

\section{Bar and spiral arm parameters}

\vspace{3 mm}
\noindent \textbf{Bar-Strength:} We decompose the face-on stellar surface density into a Fourier series (\citeauthor{Athanassoula_2002} \citeyear{Athanassoula_2002}).
\begin{equation}
    \Sigma(r,\theta)=\sum_{m=0,1,2,3...} A_{m}(r)e^{im\theta}
\end{equation}

The ratio of the amplitude of the second order and zero$^{th}$ order terms of the Fourier expansion given by

\begin{equation}
A_{2}(R)= \frac{|\Sigma_{j}m_je^{2i\theta_j}|}{\Sigma_{j}m_j}
\label{eq:bar}
\end{equation}
The bar strength is value of the peak of $A_2(R)$ ($A_2{max}$)
and phase of bar $\Phi(R)$ is given by
\begin{equation}
\Phi(R)= \frac{1}{2}{\rm arctan}\bigg[\frac{\Sigma_{j}m_j{\rm sin}(2\theta_j)}{\Sigma_{j}m_j\rm{cos}(2\theta_j) }  \bigg],
\label{eq:phase}
\end{equation}
where \( m_j \) is the mass of the jth particle, and \( \theta_j \) is the azimuthal angle. To compute \( A_2(R) \) and \( \Phi(R) \), summations are performed over all the stellar particles within an annulus of radial width \( \Delta r = 0.1 \, kpc \).
\vspace{3 mm}

\noindent\textbf{Bar-length:} Bars in the galactic disks have a constant phase angle. Hence, the phase $\Phi(R)$ should remain constant inside the bar radius, but in simulations, it deviates. Let $\sigma$ be the standard deviation of $\Phi(R)$. The bar length ($r_{bar}$) is the radius at which $\sigma$ is less than or equal to 0.1 (\citeauthor{Rosas-Guevara_2022} \citeyear{Rosas-Guevara_2022}).
\vspace{3 mm}
\\

\noindent\textbf{Bar-pattern speed:} Being a density wave, the bars rotate about the galactic centre with a constant pattern speed like a rigid body. To measure the bar pattern speed $\Omega_{\mathrm{p}}$, we use the Python code \texttt{patternSpeed.py} (\citeauthor{Dehnen2023} \citeyear{Dehnen2023}), version 0.5.3. In the code, we manually input the parameters of the bar region, such as the centre of the bar, the bar radius obtained from the Fourier analysis and measure the pattern speed corresponding to the $m=2$ mode.
\vspace{3 mm} 
 \\

\noindent\textbf{Spiral amplitude:} Here, we use the value of the peak of $A_2(R)$ at R greater than the bar radius to quantify the spiral amplitude.
\vspace{3 mm}
\\
 
\noindent\textbf{Pitch angle of spiral arms:}  We determine the \emph{pitch angle} ($\Psi$) of spiral arms using the publicly available code \textbf{P2DFFT} (parallelized two-dimensional fast Fourier transform algorithm) (\citeauthor{Mutlu_2017} \citeyear{Mutlu_2017}, \citeauthor{Hewitt_2020} \citeyear{Hewitt_2020}). In two-armed spirals, the spiral pattern is often logarithmic in nature (\citeauthor{Seigar_1998} \citeyear{Seigar_1998}). The logarithmic spiral can be written in polar coordinates $(r, \theta)$ as follows:

\begin{equation}
r(\theta) = R \cdot e^{\theta \tan \phi}
\end{equation}

where $\theta = 0$ at $r = R \in \mathbb{R}$ and $\phi$ is the pitch angle. P2DFFT decomposes galaxy images into logarithmic spirals and determines the pitch angle that maximizes the Fourier amplitudes for each harmonic mode (m). In P2DFFT, we input a FITS image of stellar surface density in face-on orientation and choose the m=2 mode in which pitch-angle is constant over the larger radius range, and measure pitch angle by averaging over this radius range.

\section{Sample} \label{TNGsamples}

The IllustrisTNG (The Next Generation) project (\citeauthor{Nelson2019b} \citeyear{Nelson2019b}, \citeauthor{Nelson2019} \citeyear{Nelson2019},  \citeauthor{Pillepich2019} \citeyear{Pillepich2019}) is a cosmological, gravitomagnetic-hydrodynamical simulation of galaxy formation, in which simulations run with the moving-mesh \textsc{AREPO} code (\citeauthor{springel_2010} \citeyear{springel_2010}). The run TNG50-1 has the highest resolution with a typical stellar particle mass of $8.5\times10^4\, M_\odot$ and a gravitational softening length of 288\,pc. This study focuses on a sample of barred-spiral galaxies from the TNG50-1 run of the IllustrisTNG suite of simulations.

\begin{figure}[H]
    \begin{center}
    \begin{tabular}{cc}
    \parbox[c]{35mm}{\centering (a)\\
    \includegraphics[width=35mm]{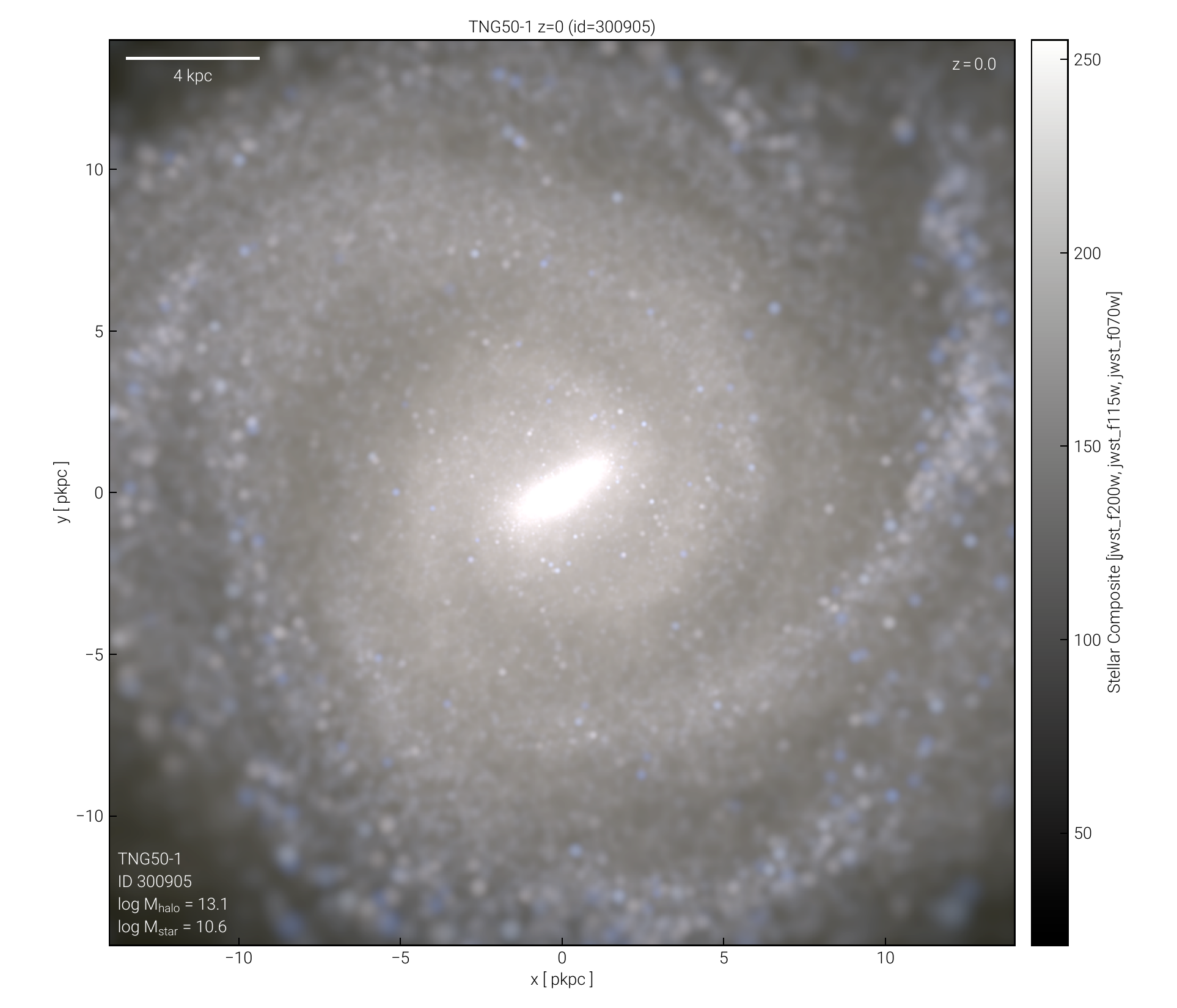}} &
    \parbox[c]{35mm}{\centering (b)\\
    \includegraphics[width=35mm]{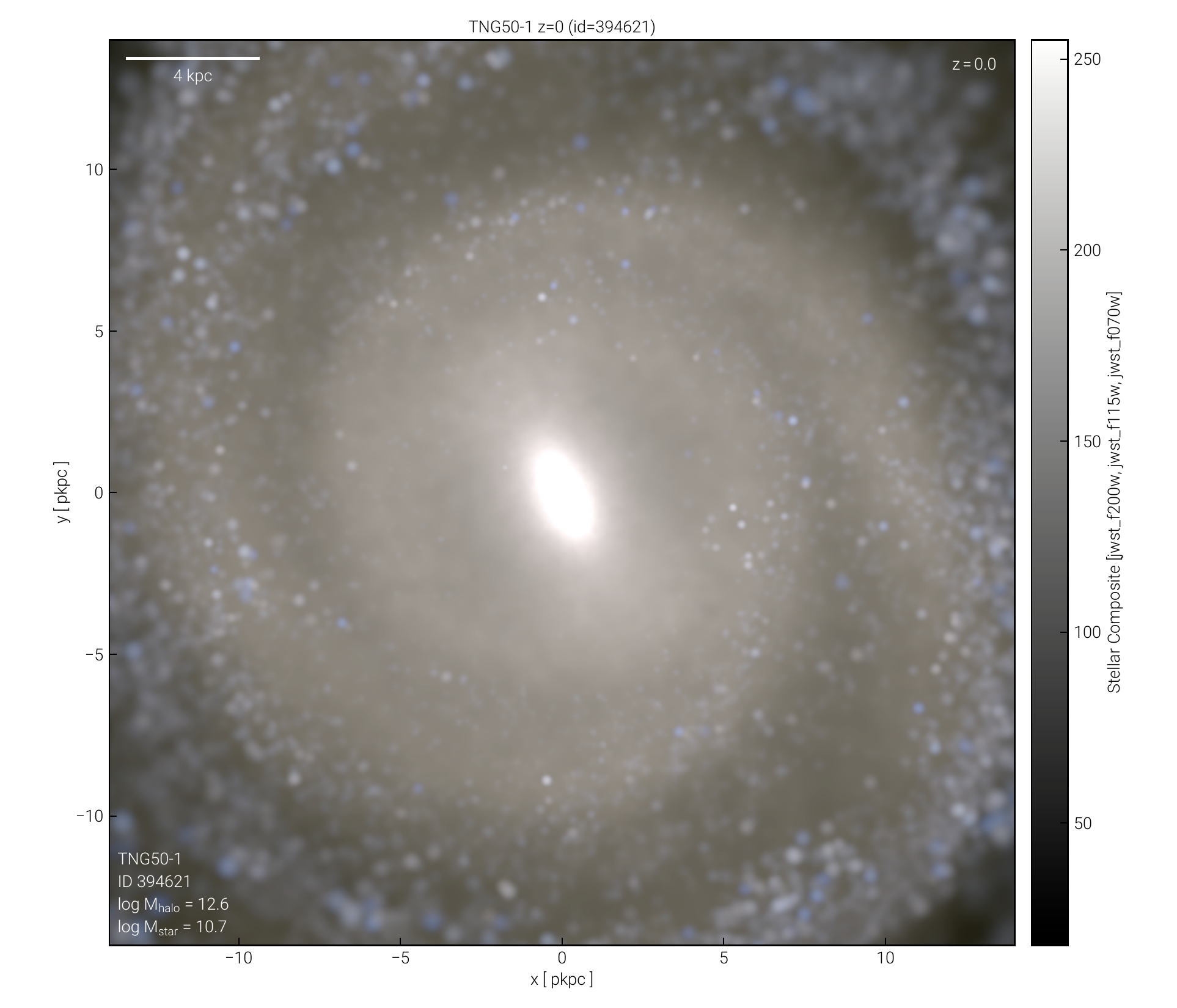}} \\
    \parbox[c]{35mm}{\centering (c)\\
    \includegraphics[width=35mm]{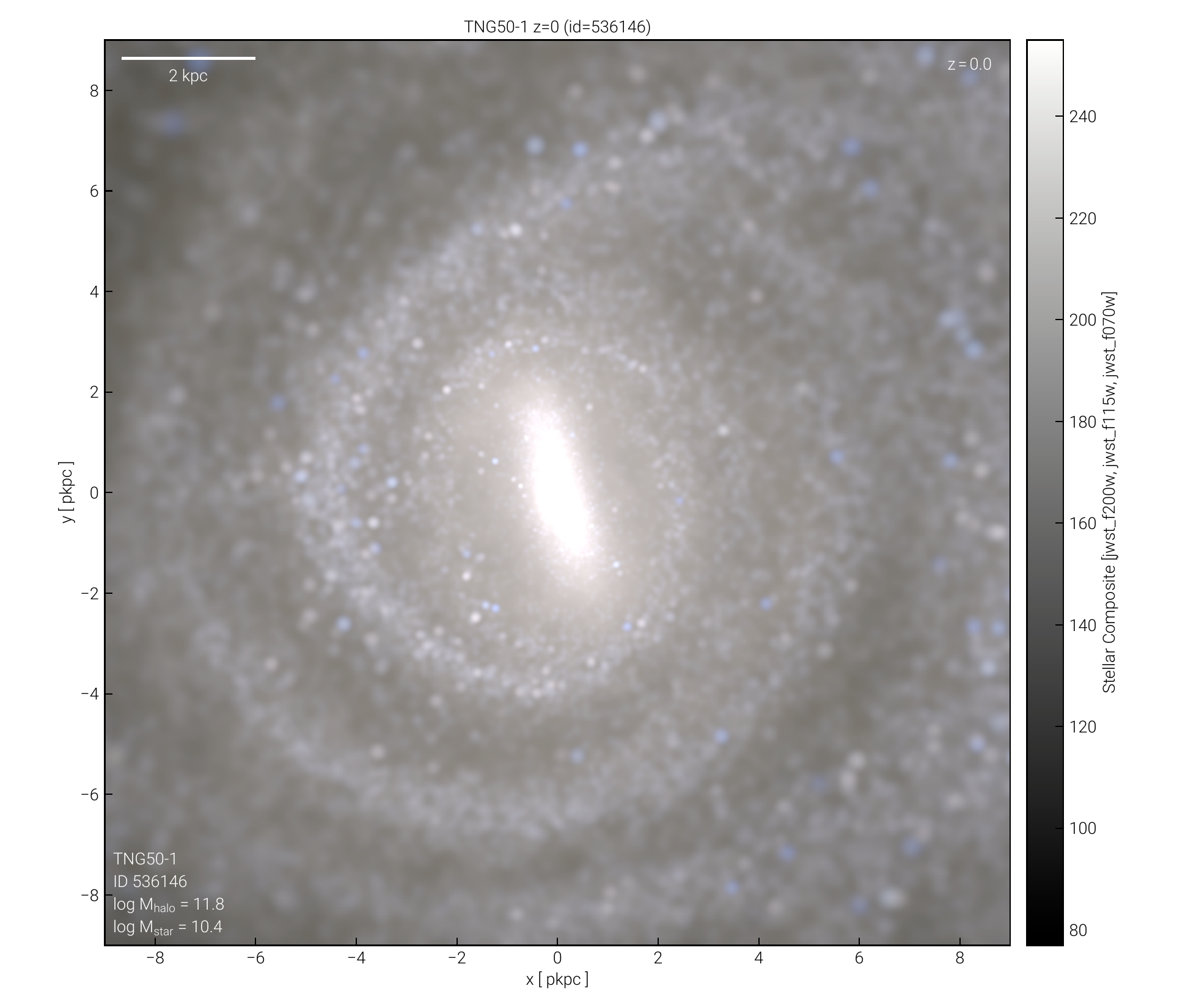}} &
    \parbox[c]{35mm}{\centering (d)\\
    \includegraphics[width=35mm]{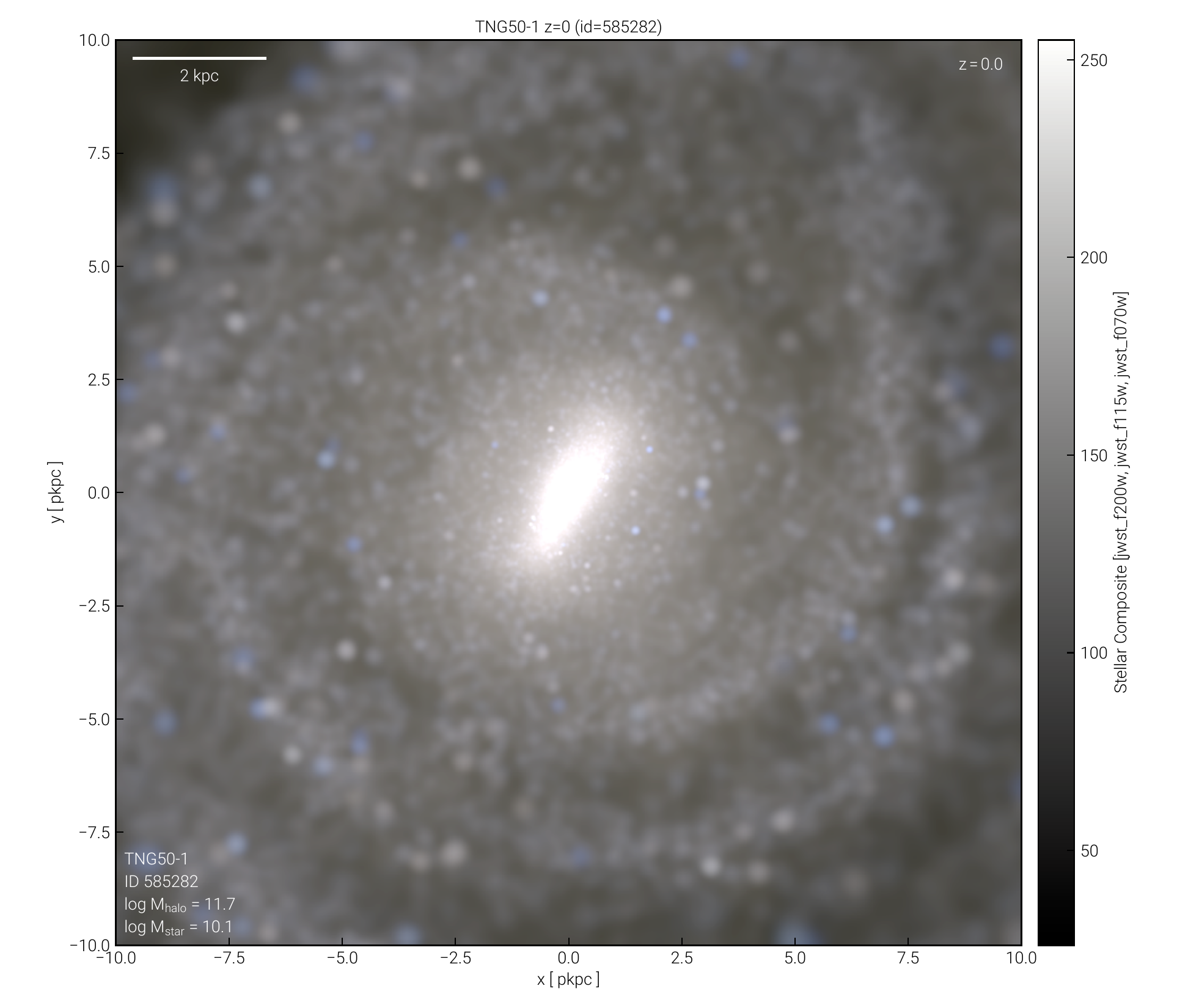}} \\
    \end{tabular}
    \end{center}
    \caption{Examples of barred-spiral galaxies at redshift $z = 0$ from the TNG50 simulation. Each panel shows composite stellar images using JWST filters [$f200w$, $f115w$, $f070w$] in a face-on orientation.}
    \label{sample_z_0}
\end{figure}

\begin{figure}[H]
    \begin{center}
    \includegraphics[scale=0.3]{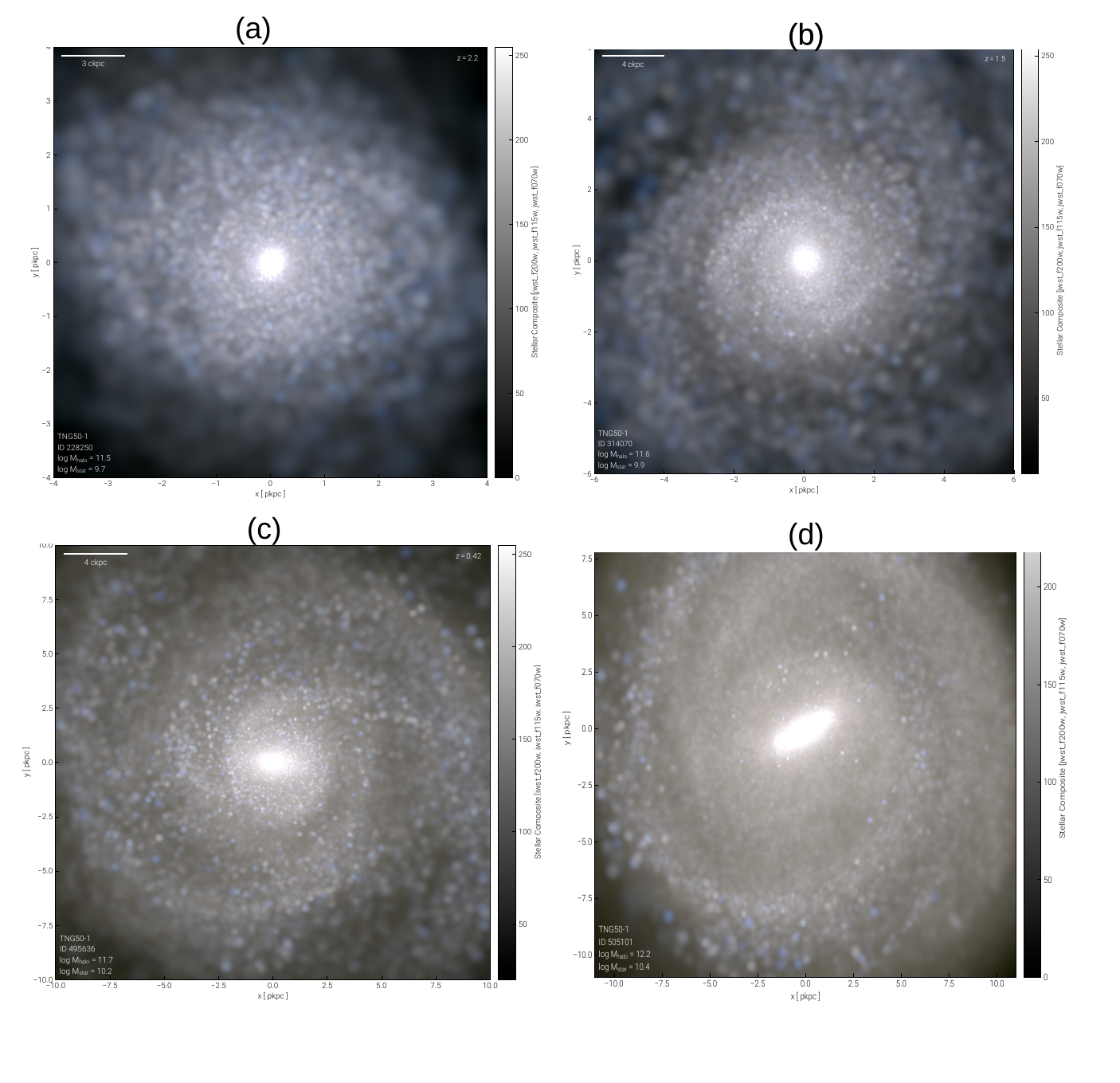}
    \end{center}
\caption{Evolution of the barred-spiral galaxy (ID 505101, z=0) in the TNG50 simulation. In this galaxy, the spiral arm is formed before the bar (spiral arm precedes bar). Each panel shows a face-on composite stellar image using JWST bands [F070W (blue), F115W (green), F200W (red)] at different redshifts (Z). (a) z=1.5, (b) z=0.42 (epoch at which bar is formed), (c) z=2.2 (epoch at which spiral arm is formed), and (d) z=0.}
\label{505101}
\end{figure}

\begin{figure}[H]
    \begin{center}
    \includegraphics[scale=0.3]{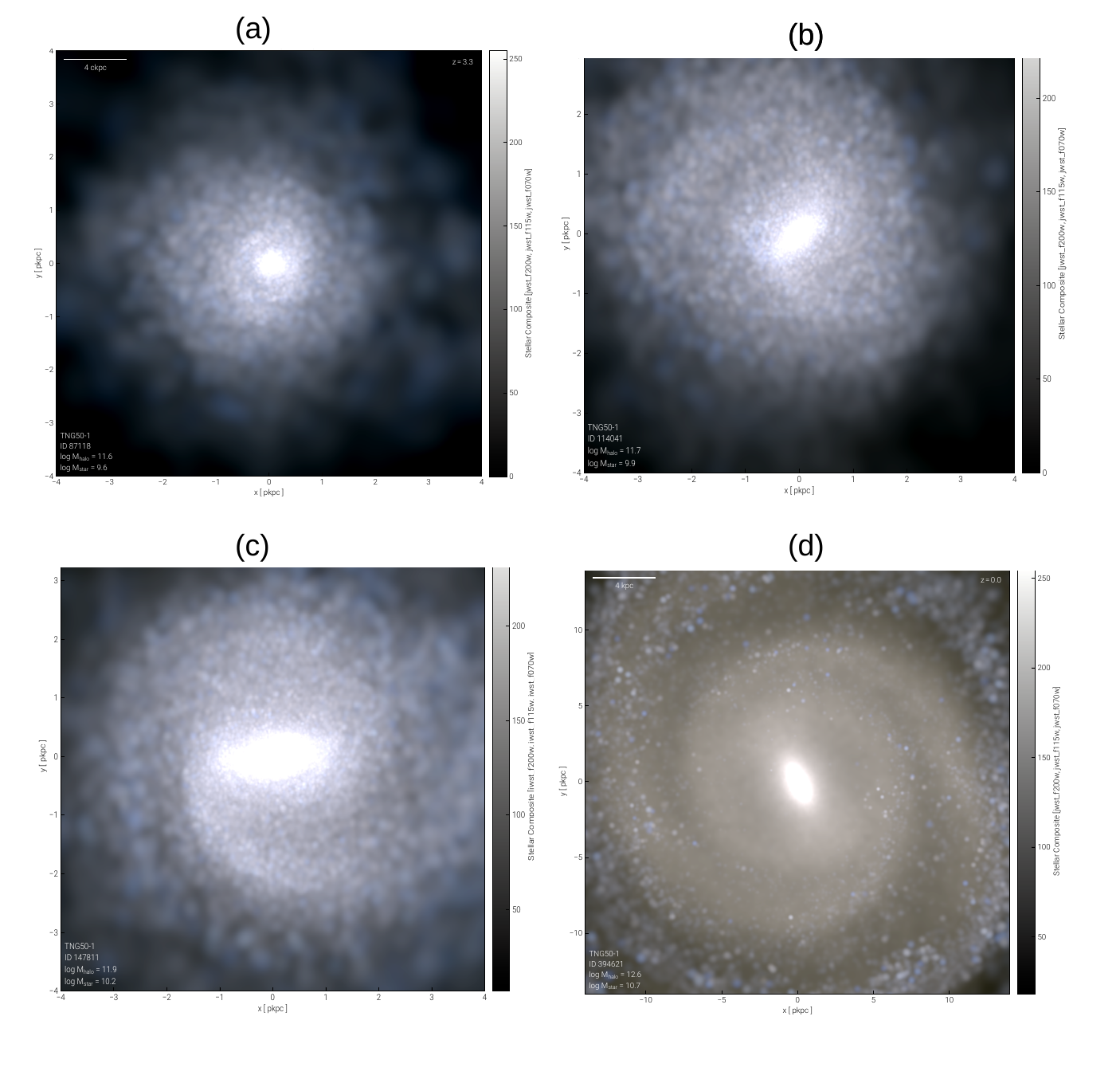}
    \end{center}
\caption{Evolution of the barred-spiral galaxy (ID 394621, z=0) in the TNG50 simulation. In this galaxy, a bar forms before the spiral arm (bar precedes spiral). Each panel shows a face-on composite stellar image using JWST bands [F070W (blue), F115W (green), F200W (red)] at different redshifts (Z). (a) z=3.3, (b) z=2.7 (epoch at which bar is formed), (c) z=2.2 (epoch at which spiral arm is formed), and (d) z=0.}
\label{394621}
\end{figure}

\begin{figure}[H]
    \centering
    \includegraphics[scale=0.3]{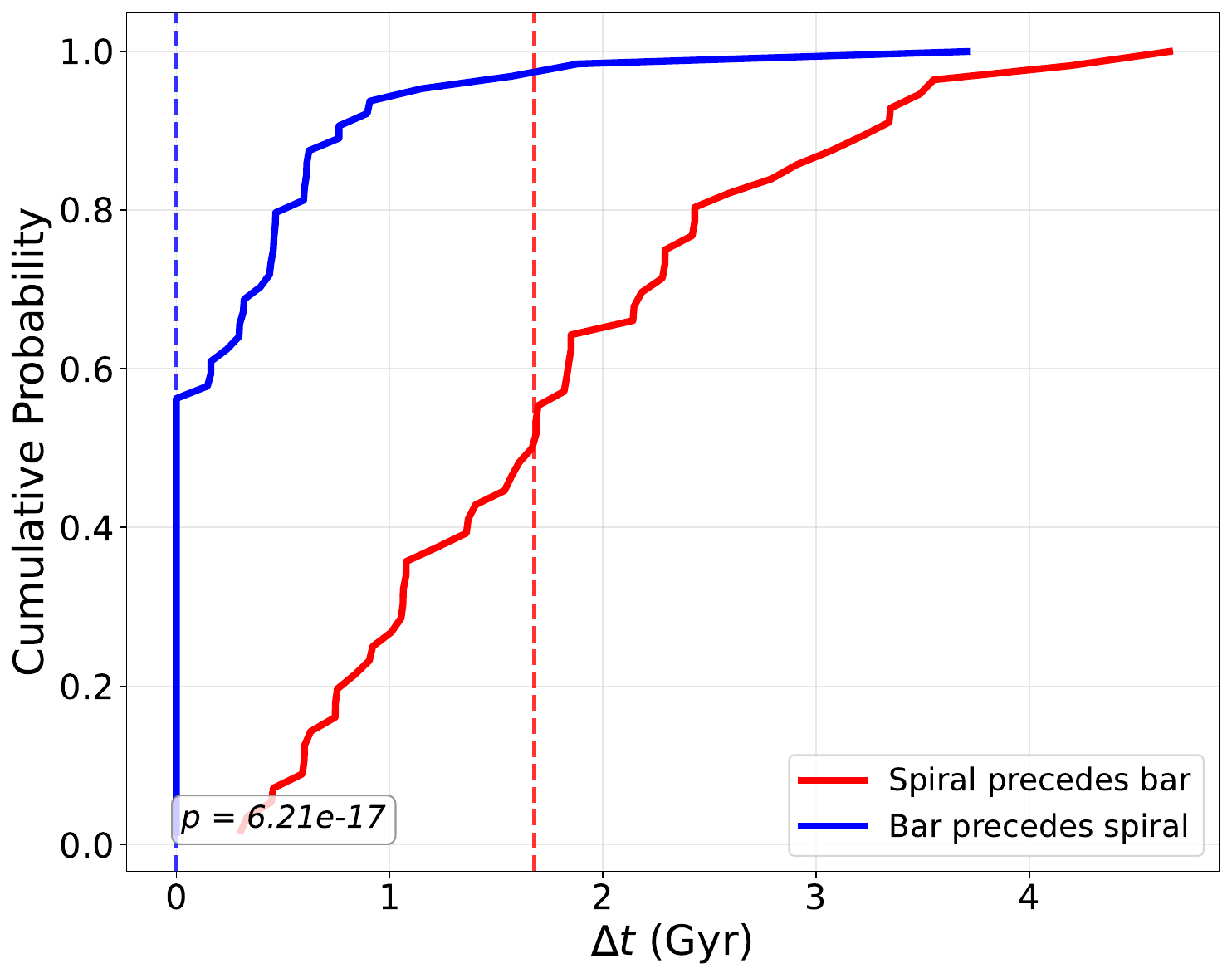}
    \caption{Cumulative distribution of time difference between appearance of bar and spiral arm in stellar disk of two galaxy samples: bar precedes spiral (orange) and spiral precedes bar (blue). Dashed lines mark the median of each distribution, and the Kolmogorov–Smirnov test p-value is indicated within the figure.}
    \label{time_diff}
\end{figure}

\begin{figure}[H]
\centering

\begin{minipage}{0.48\textwidth}
\centering
\textbf{(a)}\\
\includegraphics[width=\linewidth]{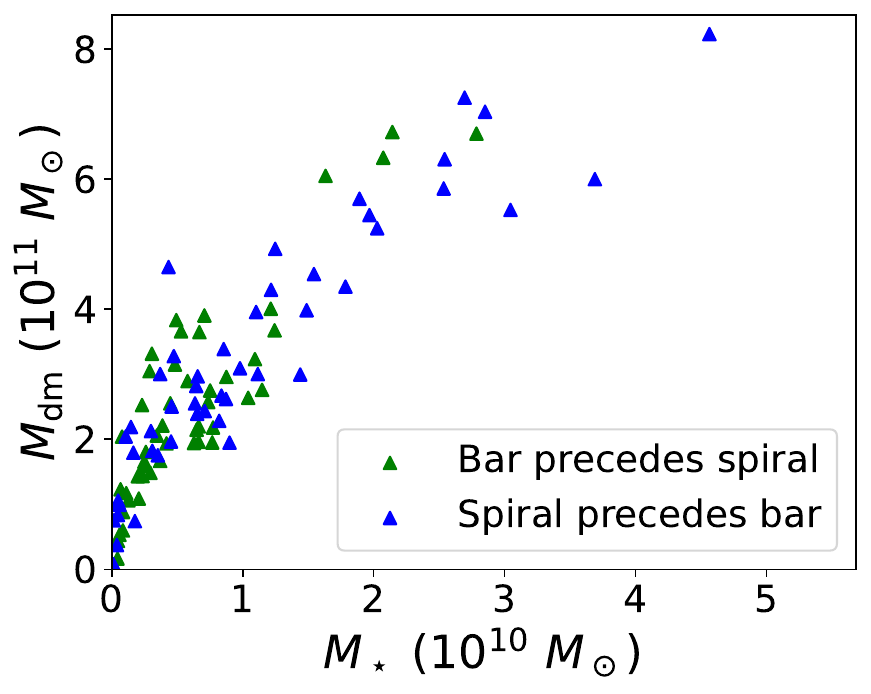}
\end{minipage}
\hfill
\begin{minipage}{0.48\textwidth}
\centering
\textbf{(b)}\\
\includegraphics[width=\linewidth]{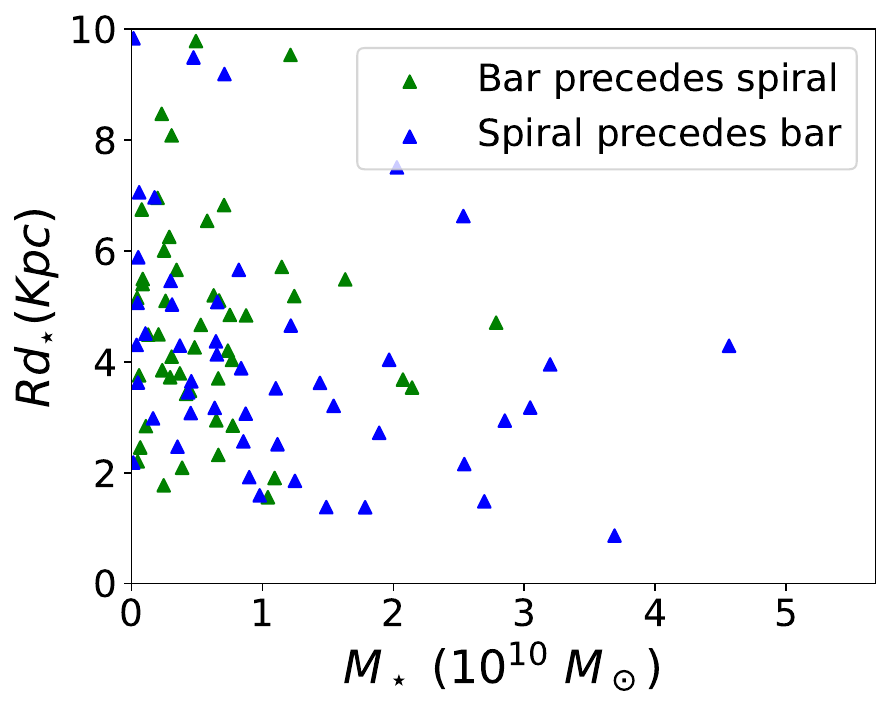}
\end{minipage}

\caption{(a) Dark matter mass and stellar mass. (b) Scale radius and stellar mass for the two groups of galaxies, bar precedes spiral (green) and spiral precedes bar (blue).}
\label{star_dm_properties}
\end{figure}

Different aspects of bars in TNG50-1 were previously studied by \citeauthor{Rosas-Guevara_2022} (\citeyear{Rosas-Guevara_2022}), \citeauthor{Frankel2022} (\citeyear{Frankel2022}), \citeauthor{IV2022} \citeyear{IV2022}), \citeauthor{Zana2022} (\citeyear{Zana2022}), \citeauthor{Ansar2023} (\citeyear{Ansar2023}), \citeauthor{Lopez2024} (\citeyear{Lopez2024}), \citeauthor{Rosas-Guevara_2024} (\citeyear{Rosas-Guevara_2024}). In this study, we focus on the barred galaxy samples identified in \cite{Zana2022}, which consist of disk galaxies with a stellar mass greater than or equal to \(10^{9} M_\odot\). We consider 750 barred galaxy samples at redshift z=0. To identify the galaxies with well-defined spiral arms, we project the stellar disk in face-on orientation by rotating the galaxy so that the z-axis aligns with the direction of the total angular momentum of the stellar component. After visually inspecting 2-dimensional plots of stellar surface density in face-on orientation, we chose a sample of barred-spiral galaxies with two-arm spirals. In Figure \ref{sample_z_0}, we show composite stellar images in a face-on orientation using JWST filters [f200w, f115w, f070w]. In the TNG simulations, as the galaxies evolve, some undergo merger events; as a result, non-axisymmetric structures get disrupted, and some evolve into lenticular galaxies. We track these galaxies and identify their progenitors within the redshift range $2 < z < 0$, removing those that exhibit large deformation in their stellar disks due to ongoing mergers/flyby events. With the above criteria imposed, we have a sample of $101$ barred galaxies at redshift $z=0$.
We classify the galaxies further into two groups based on whether the bar or spiral arm appears first in the stellar disk. Figures (\ref{505101}) and (\ref{394621}) show two of our sample galaxies and the epochs at which the bars and spiral arms formed in the stellar disks. In Table \ref{sample_t} we present the number of galaxy samples in each group and the median time difference between the formation of the bar and the spiral arm in the stellar disk. Interestingly, we note that in the first group, where the bar precedes the spirals, the spiral arms form almost immediately after the bars, whereas in the second group, where the spiral precedes the bar, the bars form about $1.7$ Gyrs after the spirals. Figure \ref{time_diff} shows the cumulative distribution of time difference between the appearance of the bar and the spiral arm in the stellar disk of two galaxy samples: bar precedes spiral (orange) and spiral precedes bar (blue). Other structural properties, such as the central mass concentration, may potentially influence this behaviour. To study this, we have analyzed dark matter mass, stellar mass, and scale radius of the stellar disk for two groups of galaxies, as shown in Figure \ref{star_dm_properties}. However, we find no significant differences between the galaxies that bar precedes spiral, and spiral precedes bar with respect to these quantities.
 The time difference in the formation of bar and spiral structures within a stellar disc may depend on the galaxy's formation history. Additionally, this difference could be influenced by one of the non-axisymmetric structures, which could be affecting the growth of the other. An early bar might trigger or enhance spiral structures, or existing spiral instabilities, delaying the subsequent formation of a bar. 
In Figure \ref{bar_parameters}, we show the evolution of bar length, bar strength and pattern speed for a barred spiral galaxy at redshift z=0 (ID: 394621).

\begin{table}[H]
\centering
\caption{Median time interval between bar and spiral arm formation for different evolutionary sequences}
\begin{tabular}{lcc}

\hline
\textbf{Galaxy sample } & \textbf{Number} & \textbf{$\Delta t$ (Gyr)} \\
\hline
Bar precedes spiral arm   & 51 & 0.00 \\
spiral arm precedes bar   & 50 & 1.7 \\
\hline
\end{tabular}
\label{sample_t}

\end{table}

\section{Theory} \label{theory}

\subsection{Mutual Information Theory (MI)}

 The field of Information Theory was established and formalized by Claude Shannon in the 1940s (\citeauthor{shannon_1948} \citeyear{shannon_1948}). A fundamental parameter in this field is information entropy (Shannon entropy), which quantifies the amount of uncertainty associated with the value of a random variable or the outcome of a random process.

Consider two parameters $X$ ($X \in \{x_{i}:i=1,....N_x\}$) and $Y$ ($Y \in \{y_{j}:j=1,....N_y\}$) with $N_x$ and $N_y$ data points respectively (Here $N_x = N_y$ and are number of galaxies in our sample). Let X be a parameter of the bar, and Y be a parameter of the spiral arm. We divide the available values of X and Y into $N_d$ bins. The probability of finding a randomly selected parameter value in the $i^{th}$ bin is 
\begin{eqnarray}
    p(X_i)=\frac{N_i}{N}
\end{eqnarray}

where $N_i$ is the number of observables of X in the $i^{th}$ bin and $N$ is the total number of galaxies. The information entropy (\citeauthor{shannon_1948} \citeyear{shannon_1948}) associated with the random variable $X$ is given by
\begin{eqnarray}
    H(X)& = &-\sum_{i=1}^{N_d} p(X_i) \log p(X_i) 
  \label{eqn:Hx}
\end{eqnarray}

 Here, the logarithms are to base 2.

Similarly, the Shannon entropy for variable Y is given as,

\begin{eqnarray}
    H(Y)& = &-\sum_{i=1}^{N_d} p(Y_i) \log p(Y_i) 
  \label{eqn:HY}
\end{eqnarray}

The joint entropy $H(X,Y)$ on the other hand, is given as
\begin{eqnarray}
H(X,Y) & = & - \sum \limits_{i=1}^{N_d}\sum \limits_{j=1}^{N_d} \, p(X_i,Y_j) \, \log \, p(X_i,X_j)
\label{eq:joint}
\end{eqnarray}

The Mutual Information between the bar and spiral parameters ($X$ and $Y$) is defined as 

\begin{eqnarray}
I(X;Y) & = & H(X)+H(Y)-H(X,Y),
\end{eqnarray}
$I(X; Y)=0$ if X and Y are statistically independent and $I(X; Y)=H(X)$ if X is completely determined by Y. Normalised Mutual Information is given by
\begin{eqnarray}
NMI(X;Y) & = \frac{2*I(X;Y)}{H(X)+H(Y)},
\label{Neq_mutual}
\end{eqnarray}

Normalized Mutual Information has a value ranging between $[0,1]$. We note, $I(X;Y)$ and $NMI(X;Y)$ is
symmetric under the exchange of $I$ and $J$ and therefore does not contain any
directional sense. MI measures the amount of shared information between two mutually associated variables. Hence, it can also be used to examine the mutual exclusiveness of two quantities (\citeauthor{Thommas_2006} \citeyear{Thommas_2006}). Mutual Information (MI) is a model-free measure that quantifies the statistical dependence between variables, unlike traditional correlation measurements that focus on linear relationships. For instance, Pearson's correlation coefficient approaches zero when the relationship between two variables is nonlinear. While traditional nonlinear regression methods can model these relationships, they often require a predetermined functional form and can be challenging to analyze when the underlying connections are unknown or highly complex. In contrast, MI measures how much knowing one variable reduces the uncertainty about another. Consequently, it captures a wide range of statistical dependencies, including linear, nonlinear, and complex relations that are difficult to parameterize.
A high value of MI indicates that two random variables are dependent, and a zero value of MI suggests that the two random variables are independent.

\subsection{Transfer Entropy}

The Shannon entropy and its generalization, the MI, are
properties of the static probability distributions, while the dynamics of the
processes are contained in the transition probabilities. We measure the Transfer Entropy (TE) to study the dynamics of shared information between parameters. TE measures determine the direction and quantify the information transfer between two processes (\citeauthor{Schreiber_2000} \citeyear{Schreiber_2000}). Consider the time series data of a bar parameter I ($I \in \{i_{n}:n=1,....N\}$) and a spiral parameter J ($J \in \{j_{n}:n=1,....N\}$). The TE is given by
\begin{eqnarray}
    T_{J\to I} = \sum\,
         p(i_{n+1},i^{(k)}_n,j^{(l)}_n)\, \log \,
         {\frac{p(i_{n+1}|i^{(k)}_n,j^{(l)}_n)} {p(i_{n+1}|i^{(k)}_n)}}
         \label{eq:transfer}
\end{eqnarray}

 where $p(i_{n+1}|i^{(k)}_n)=p(i^{(k+1)}_{n+1})/p(i^{(k)}_n)$ is the conditional probability. $k$ denotes the number of past timesteps of the bar parameter $I$ included in predicting its future value,

\begin{equation}
i_n^{(k)} = (i_n, i_{n-1}, \dots, i_{n-k+1}),
\end{equation}

while $l$ denotes the number of past timesteps of the spiral parameter $J$ included additionally in the prediction.

\begin{equation}
j_n^{(l)} = (j_n, j_{n-1}, \dots, j_{n-l+1}).
\end{equation}

In the present work, we adopt the first-order choice.

\begin{equation}
k = l = 1,
\end{equation}

which corresponds to using only the immediately preceding values $(i_n, j_n)$ to estimate the next value $i_{n+1}$. 
$T_{J\to I}$ is now explicitly non-symmetric since it measures the degree of dependence of $I$ on $J$ and not vice versa. Followed by \citeauthor{NTE1} \citeyear{NTE1}, \citeauthor{NTE2} \citeyear{NTE2} Normalised Transfer Entropy is given by
\begin{equation}
\mathrm{NTE}_{J \rightarrow I} =
\frac{\mathrm{TE}_{J \rightarrow I} - \left\langle \mathrm{TE}_{J_{\mathrm{shuffle}} \rightarrow I} \right\rangle}
{H(I_{n+1} \mid I_n)} ,
\label{NTE}
\end{equation}

where
 $\left\langle \mathrm{TE}_{J_{\mathrm{shuffle}} \rightarrow I} \right\rangle$ represents the average shuffled transfer entropy from $J$ to $I$, computed by randomly permuting the time series of $J$, recalculating $\mathrm{TE}_{J_{\mathrm{shuffle}} \rightarrow I}$ for each realization, repeating the procedure 100 times, and taking the mean of the resulting transfer entropy values. ,
and $H(I_{n+1} \mid I_n)$ is the conditional entropy of $I$ at time $n+1$ given its value at time $n$, given by.

\begin{equation}
H(I_{n+1} \mid I_n) =
- \sum_{I_{n+1},\, I_n}
p(I_{n+1}, I_n)
\log \left(
\frac{p(I_{n+1}, I_n)}{p(I_n)}
\right).
\end{equation}
 Now $NTE_{J\to I}$ is in the range [0,1] (\citeauthor{NTE1}, \citeyear{NTE1}).  $NTE_{J\to I}=0$ when J transfers no information to I, and is 1 when J transfers maximal information to I.

We consider $101$ barred spiral galaxies at a redshift of $z=0$ mentioned in section $\S $ \ref{TNGsamples}. We track each galaxy and identify its progenitors within the redshift range $2 < z < 0$, and obtain time-series data for the bar and spiral-arm parameters. For the time series data of a bar parameter I ($I \in \{i_{n}:n=1,....N\}$) and a spiral parameter J ($J \in \{j_{n}:n=1,....N\}$), the transfer entropy implemented using the Python implementation by Sebastiano Bontorin (\texttt{transfer\_entropy}) (\citeauthor{BontorinTE} \citeyear{BontorinTE}), which estimates the transfer entropy introduced by \citeauthor{Schreiber_2000} (\citeyear{Schreiber_2000}). The probability distributions required for the Transfer Entropy calculation are estimated using histogram discretization of the time series I and J. The number of bins for each variable was determined using the  {Freedman--Diaconis rule}    (\citeauthor{Freedman1981} \citeyear{Freedman1981}), which defines the optimal histogram bin width as

\begin{equation}
h = \frac{2\,\mathrm{IQR}}{n^{1/3}},
\end{equation}

where $\mathrm{IQR}$ is the interquartile range of the data and $n$ is the number of samples in the time series. 

For J the number of bins is then obtained as

\begin{equation}
N_{\mathrm{bins}} = \frac{\max(J) - \min(J)}{h}.
\end{equation}

This adaptive binning method accounts for both the spread of the data and the sample size and is commonly used to estimate probability densities from empirical data. We adopt this rule as an objective, widely used, data-driven method for histogram construction, thereby avoiding arbitrary bin selection.
The same procedure is applied independently to both time series $I$ and $J$ to construct the joint and marginal probability distributions required for the entropy calculations. Followed by TE estimation, we calculated Normalized transfer entropy using equation \ref{NTE}. We note that the estimated Transfer Entropy values are sensitive to the number of bins.

While classical implementations of transfer entropy (TE) often assume stationarity for practical estimation of probability distributions, the definition of TE itself does not require stationarity. Transfer entropy is fundamentally a conditional mutual information measure, and is therefore valid for arbitrary stochastic processes, including non-stationary ones. We employ a time-resolved (or windowed) estimation approach, in which TE is computed within short temporal windows. Within each window, we make sure that each window has at least 40 data points and in windows, changes in the mean and variance remained below 5$\%$ and $10\%$, respectively and obtain the NTE values {equation \ref{NTE}} for both bar-to-spiral and spiral-to-bar for all bar/spiral arm parameter pairs. The final TE values are obtained by taking the average of the TE computed over all temporal windows.

\subsection{Information Flow}
The concept of the Liang information flow rate (IFR) is based on information entropy and the theory of dynamical systems (\citeauthor{Liang_2008} \citeyear{Liang_2008}, \citeauthor{liang_2013} \citeyear{liang_2013}, \citeauthor{liang_2014} \citeyear{liang_2014}, \citeauthor{liang_2015} \citeyear{liang_2015}). Like TE, IFR also determines the direction and quantifies the information transfer between two processes. Consider the time series data of bar parameter I ($I \in \{i_{n}:n=1,....N\}$) and spiral parameter J ($J \in \{j_{n}:n=1,....N\}$). The sample covariance between $I$ and $J$ is defined as

\begin{eqnarray}
    C_{IJ} = \overline{(I - \overline{I})(J - \overline{J})}    
\end{eqnarray}
 and covariance between $J$ and $\dot I$ is defined as,
 
 \begin{eqnarray}
    C_{J,dI} = \overline{(J - \overline{J})(\dot{I} - \overline{\dot{I}})}
\end{eqnarray}
 
 Here the overbar denotes the sample mean, $\dot I$ is the difference approximation of $dI/dt$ using the Euler forward scheme
\begin{eqnarray}
    \dot I(n) = \frac {I(n+k) - I(n)} {k \Delta t}.
\end{eqnarray}

Here, $k$ is set to 1. The time series is sampled at regular intervals

We use the Liang-Kleeman coefficient (\citeauthor{liang_kleeman_2005} \citeyear{liang_kleeman_2005}, \citeauthor{liang_2014} \citeyear{liang_2014})  to quantify the directional transfer of information across time series. Following Equation (10) of \citeauthor{liang_2014} (\citeyear{liang_2014}), the IFR is given by 

\begin{eqnarray}	
	T_{J \to I} = \frac{C_{II} C_{IJ} C_{J,dI} - C_{IJ}^2 C_{I,dI}} 
			     {C_{II}^2 C_{JJ} - C_{II} C_{IJ}^2},
    \label{eq:TJI_est}
\end{eqnarray}

 This formula uses only sample covariances, which makes it easy to compute. If \(C_{IJ} = 0\), then \(T_{J\to I} = 0\) However, if \(T_{J\to I} = 0\), it doesn’t mean that \(C_{IJ}\) has to be zero. Causation leads to correlation, but correlation does not necessarily mean there is causation.

The overall rate of entropy change in the receiver(I) is influenced not just by the information transfer from the transmitter(J), characterized by the rate \( T_{J \rightarrow I} \), but also by the intrinsic entropy rate (  \( \frac{dH_I^*}{dt} \)) and the noise-induced entropy rate \( \frac{dH_I^s}{dt} \)( \citeauthor{liang_2015} \citeyear{liang_2015}). 

\begin{subequations}
\begin{align}
\frac{d H_{I}^*}{dt} &= p_{JI}  \\[1ex]
\frac{d H_{I}^s}{dt} &= \frac{\Delta t}{2 C_{II}} \Big(
    C_{dI,dI}
    + p_{JI}^2 C_{II}
    + q_{JI}^2 C_{JJ} \notag \\
    &\quad - 2 p_{JI} C_{dI,I}
    - 2 q_{JI} C_{dI,J}
    + 2 p_{JI} q_{JI} C_{IJ}
\Big) 
\end{align}
\end{subequations}
Where
\begin{subequations}
\begin{align}
p_{JI} &= \frac{{C}_{JJ} C_{I,dI} - C_{IJ} {C}_{J,dI} }{ C_{JJ} C_{II} - C_{IJ}^2 }  \\
q_{JI} &= \frac{ - {C}_{IJ} C_{I,dI} + {C}_{II} C_{J,dI} }{ C_{JJ} C_{II} - C_{IJ}^2 }, 
\end{align}
\end{subequations}

Following \citeauthor{liang_2015} \citeyear{liang_2015}, the normalisation factor is given by
\begin{eqnarray}	
	          Z  \equiv |{T_{J\to I}}| + 
				|{\frac{dH_I^*}{dt}}| +
				|{\frac {dH_I^{noise}}{dt}}|.
        \label{eq:normalizer}
\end{eqnarray}

Thus, the normalized IFR is 
\begin{eqnarray}
	\tau_{J\to I} = T_{J\to I} / Z.
    \label{NIF}
\end{eqnarray}

$\tau_{J \rightarrow I}$ measures the percentage of the total entropy rate of change for $I$ which is due to its interaction with $J$ ({\citeauthor{liang_2015} \citeyear{liang_2015}). The normalized IFR values lie within the range 0 to 1.

We consider $101$ barred spiral galaxies at a redshift of $z=0$ mentioned in section $\S $ \ref{TNGsamples}. We track each galaxy and identify its progenitors within the redshift range $2 < z < 0$, and obtain time-series data for the bar (I) and spiral-arm parameters (J). We then calculate the normalized IFR values using the publicly available Python package \texttt{Pyleoclim} (\texttt{pyleoclim.utils.causality.liang\_causality}; \citeauthor{Pyleoclim} \citeyear{Pyleoclim})  for both bar-to-spiral and spiral-to-bar for all bar/spiral arm parameter pairs, for both samples of galaxies mentioned in section $\S $ \ref{TNGsamples}.

\begin{figure*}
    \centering
    \begin{tabular}{ccc}
        \parbox[c]{0.3\textwidth}{\centering \textbf{(a)}\\
        \includegraphics[width=\linewidth]{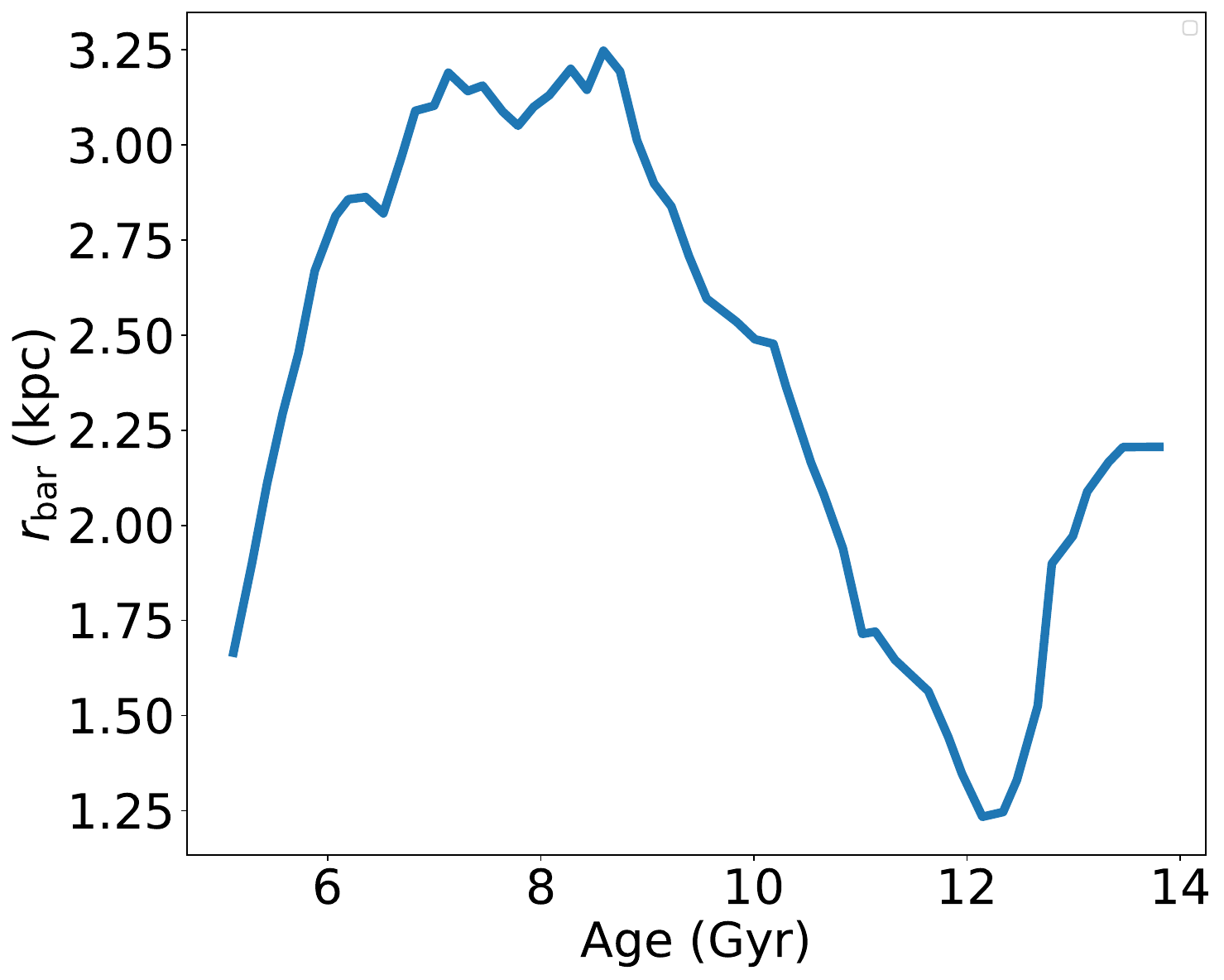}} &
        \parbox[c]{0.3\textwidth}{\centering \textbf{(b)}\\
        \includegraphics[width=\linewidth]{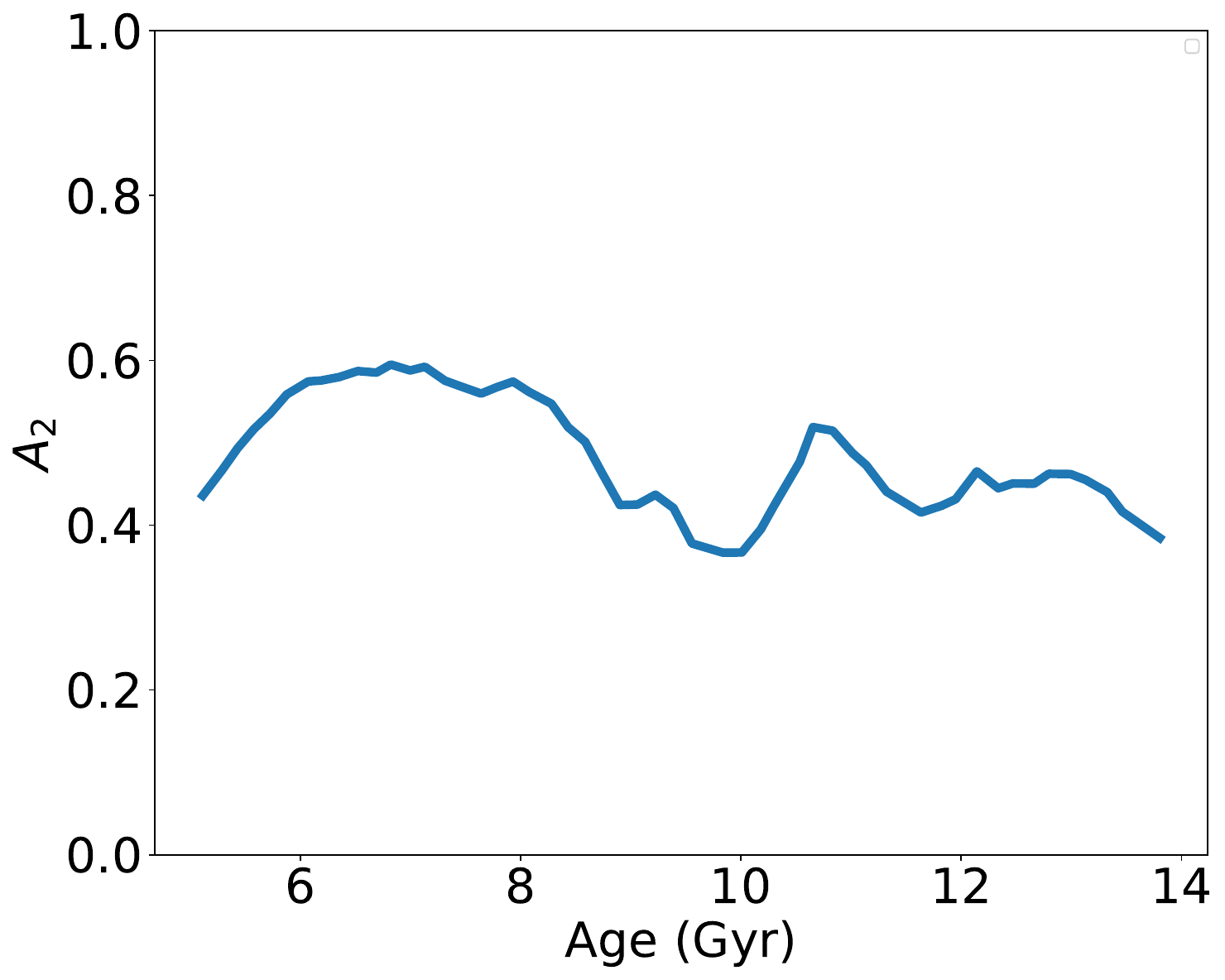}} &
        \parbox[c]{0.3\textwidth}{\centering \textbf{(c)}\\
        \includegraphics[width=\linewidth]{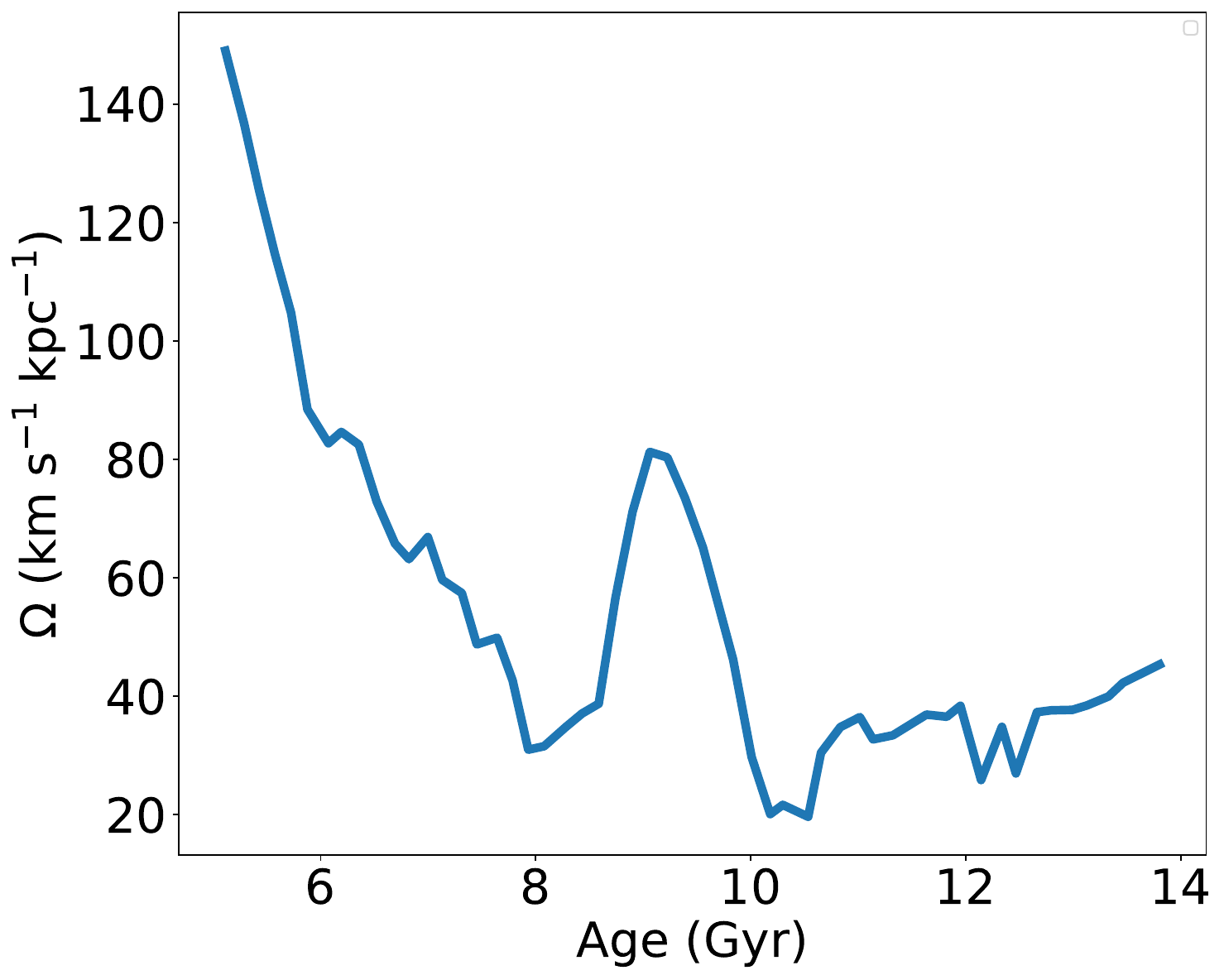}} \\
    \end{tabular}
    \caption{Evolution of (a) bar length $r_{\mathrm{bar}}$, (b) bar strength $A_{2,\mathrm{bar}}$, and (c) pattern speed $\Omega$ with time for the barred spiral galaxy (ID: 394621) at redshift $z=0$.}
    \label{bar_parameters}
\end{figure*}

\begin{figure*}
    \begin{center}
    \includegraphics[scale=0.75]{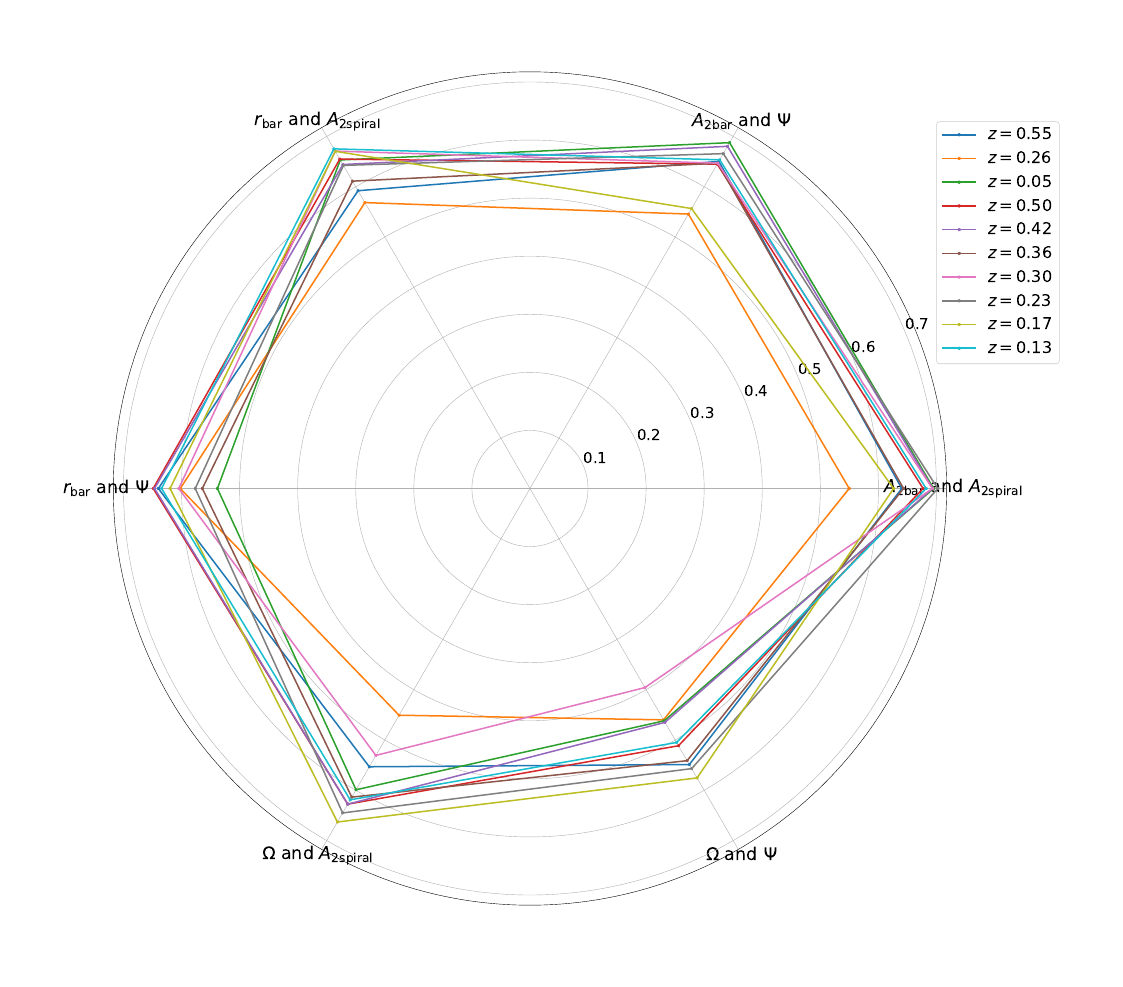}
    \end{center}
\caption{Radar plot: Comparison of Mutual Information between different pairs of structural and kinematic parameters of the bar and its associated spiral arm at different redshifts. Spikes in this radar plot correspond to Mutual Information (MI) values for different parameter pairs. The grey scale indicates the values of the Mutual Information.}
\label{Radar_MI}
\end{figure*}

\begin{figure*}
\centering
\includegraphics[width=\textwidth]{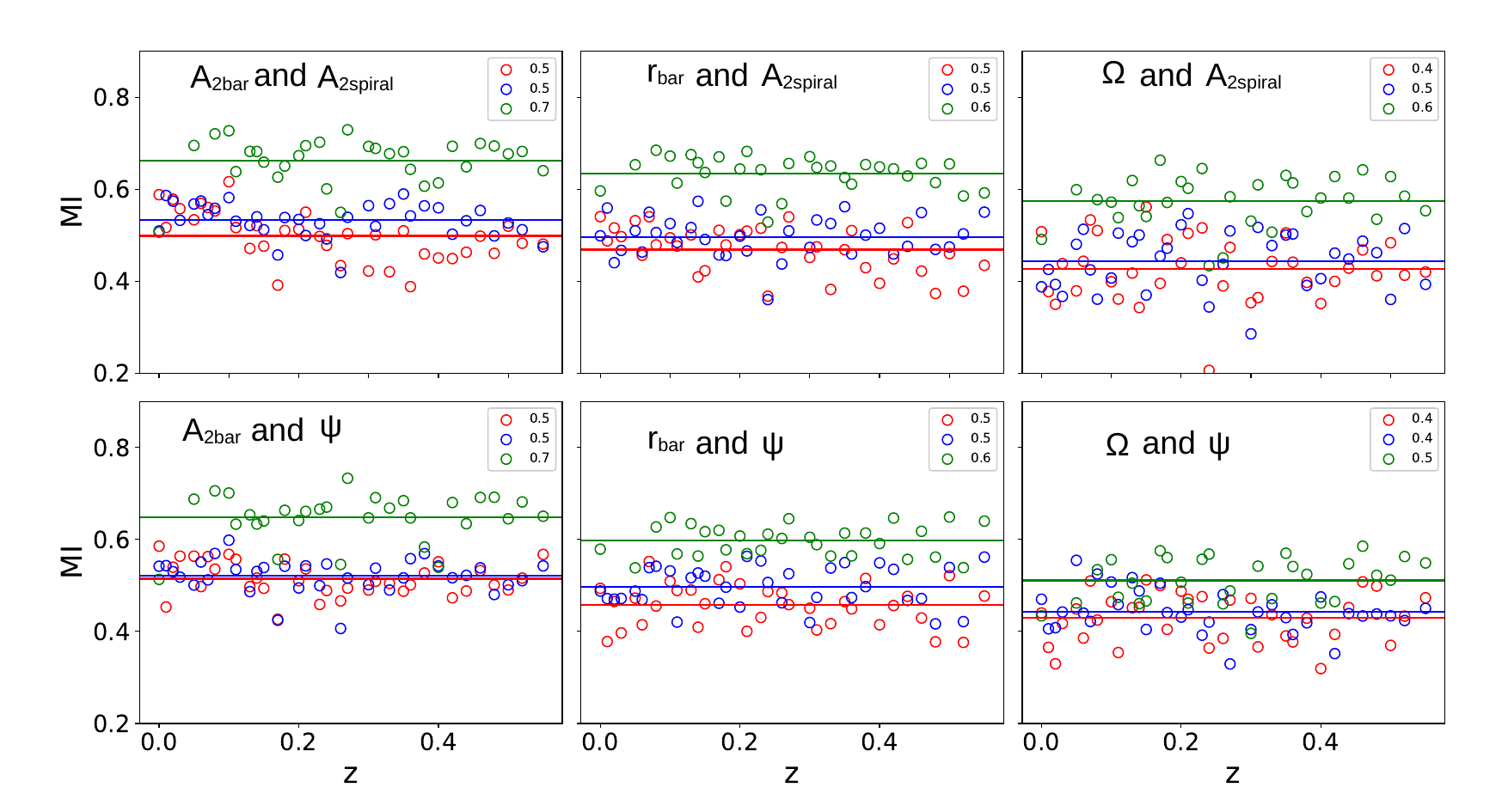}
\caption{Mutual Information between bar and spiral arm parameters for the total sample [green] and for two groups of galaxy samples: Bar precedes spiral [red], spiral precedes bar [blue]. Top Panel: Bar strength $A_{2bar}$ and spiral arm strength $A_{2spiral}$, Bar length $r_{bar}$ and spiral arm strength $A_{2spiral}$, and Bar pattern speed $\Omega$ and spiral arm strength $A_{2spiral}$]. Bottom Panel: Bar strength $A_{2bar}$ and spiral arm pitch angle $\Psi$, Bar length $r_{bar}$ and spiral arm pitch angle $\Psi$, and Bar pattern speed $\Omega$ and spiral arm pitch angle $\Psi$. In each panel, we show the average Mutual Information values by horizontal lines for the respective samples.}
\label{Total_MI}
\end{figure*}

\section{Results \& discussion}

\subsection{Mutual Information}

We calculate the value of MI to quantify the strength of the association between a pair of structural/kinematic parameters: one corresponding to the bar, and the other to the spiral arm. We consider a sample of 101 barred--spiral galaxies identified at redshift $z = 0$, as described in Section~\ref{TNGsamples}. We track each galaxy and identify its progenitors within the redshift range $0.55 < z < 0$. For this sample, we compute the mutual information (MI) between bar and spiral arm parameter pairs at each redshift. The MI values for all parameter pairs are normalized to the range $[0,1]$ using Equation~(\ref{Neq_mutual}). All parameter pairs exhibit MI values ranging from $0.4$ to $0.8$, indicating a significant degree of association between the bar and spiral arms (\citeauthor{Thommas_2006} \citeyear{Thommas_2006}), thus confirming their co-evolution. In Figure \ref{Radar_MI}, we present the MI values of different pairs of parameters in a Radar Plot. The spikes in this plot correspond to their MI values, with MI at different redshifts given in different colours. We observe that the MI values do not show significant variation with respect to redshift for any of the above parameter pairs.

We next compare the MI values for both groups of galaxies: one in which the bar precedes the spiral, and the other in which the spiral precedes the bar. In Figure \ref{Total_MI}, we show the time evolution of MI between bar and spiral arm parameters with redshift for the total sample [green] as well as for the two groups of galaxy samples: Bar precedes spiral [red], spiral precedes bar [blue]. The top (bottom) panel shows the MI between spiral arm strength $A_{2spiral}$ (spiral arm pitch angle $\Psi$) and bar parameters, i.e., bar strength $A_{2bar}$, bar length $r_{bar}$, bar pattern speed $\Omega$) and spiral arm strength $A_{2spiral}$, respectively.  In each panel, we indicate the average MI value between each pair of parameters for each case with a horizontal line. For the total galaxy sample,  the parameter pairs, $A_{2bar}$-$A_{2spiral}$, $A_{2bar}- \Psi$, $r_{bar}-A_{2spiral}$, $r_{bar}- \Psi$, $\Omega - A_{2spiral}$ exhibits an average MI value of 0.6 and the parameter pair  $\Omega - \Psi$ shows an average MI value of 0.5. \emph{We note, for both the galaxy samples (bar precedes spiral and spiral precedes bar), the mean MI is between a common range of $0.4-0.5$, revealing a strong enough and comparable degree of co-evolution of the bar and the spiral arm}. 
 Our results align with previous studies that establish the co-evolution of bars and spiral arms in galaxies. Numerical modelling by \citeauthor{schwarz_1984} (\citeyear{schwarz_1984}) shows a positive correlation between the pitch angle of spiral arms and the strength of bars. Observational evidence from real galaxies further supports this association. Using Ks-band observations from the William Herschel Telescope for a sample of 15 nearby spiral galaxies, \citeauthor{Block_2004} (\citeyear{Block_2004}) and \citeauthor{Buta_2009} (\citeyear{Buta_2009}) found that stronger bars are consistently linked to more prominent spiral arms. When bars and spirals emerge from the same instability, bars exert an influence on the spirals. Additionally, an analysis of barred-spiral galaxies from the Spitzer Survey by \citeauthor{refId0} (\citeyear{refId0}) reveals a moderate correlation between bar strength and spiral amplitude. Moreover, \citeauthor{Elmegreen_2011} (\citeyear{Elmegreen_2011}) demonstrates that longer bars with greater m=2 modes are often associated with stronger spiral structures. Also discuss the possibility that grand-design spiral arms in galaxies are formed from spiral density waves driven by bars. Since our study applies primarily to barred-spiral galaxies that simultaneously exhibit a well-defined bar and a two-armed spiral pattern in the stellar disc, and selection excludes flocculent, multi-armed, and unbarred galaxies, the extrapolation to the broader population of observed disk galaxies should be made with caution. 

However, it is important to note that MI is a static measure; it does not capture dynamic evolution or indicate a causal relationship between two random variables.

\subsection{Transfer Entropy \& Information Flow}

\begin{figure*}[hbt]
    \begin{center}
    \includegraphics[scale=0.45]{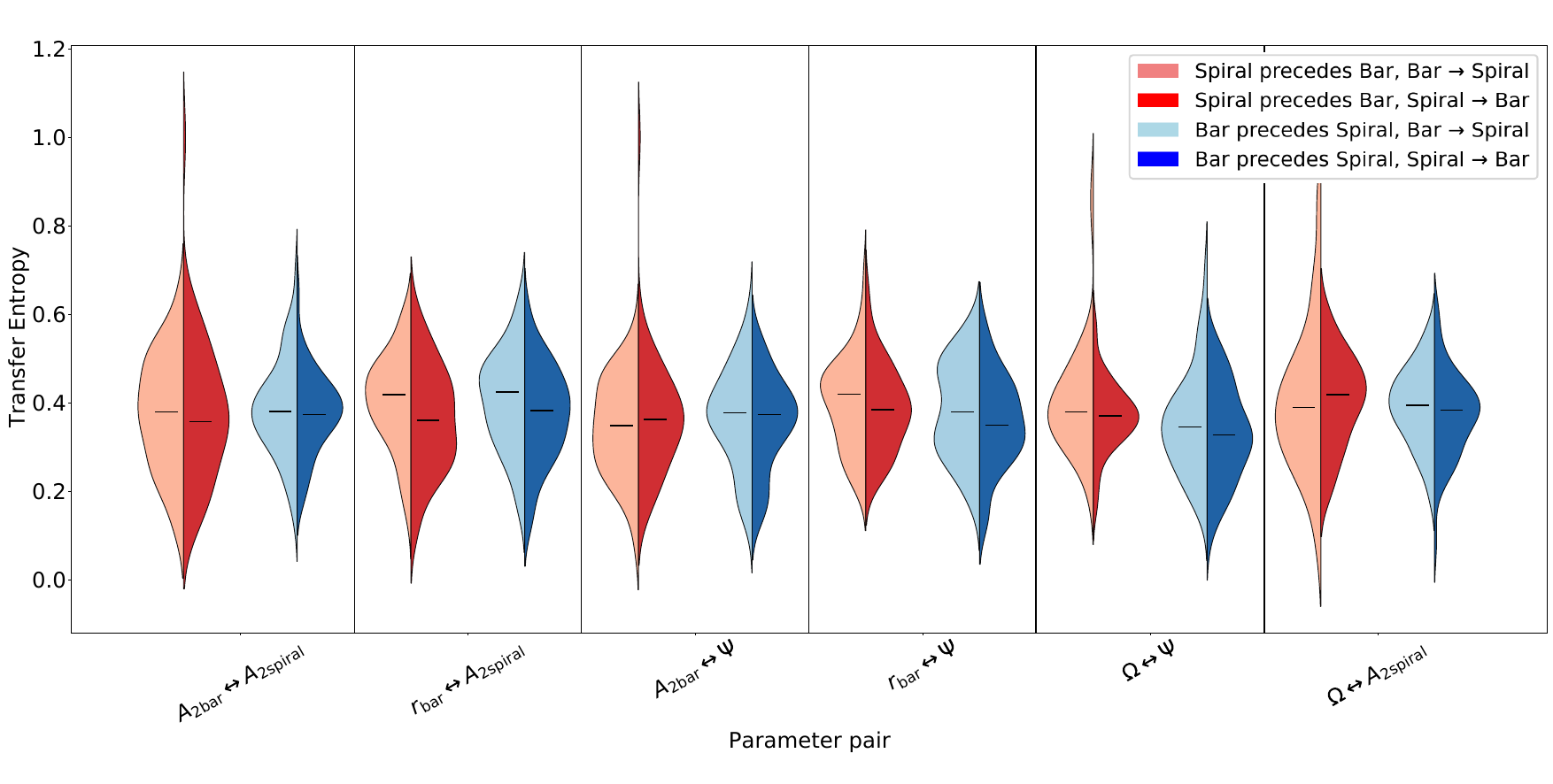}
    \end{center}
    \caption{Violin plots of Transfer Entropy values for different parameter pairs. The two sides of the violin illustrate the distribution of Transfer Entropy in two opposite directions: Transfer Entropy from spiral to bar ($T_{\text{spiral\ to\ bar}}$) (darker shade) and Transfer Entropy from bar to spiral ($T_{\text{bar\ to\ spiral}}$) (lighter shade). Red shades represent galaxies where spiral precedes bar, while blue shades indicate galaxies where bar precedes spiral. Black horizontal lines inside the violins denote the median value of Transfer Entropy for each direction. Left to right panels correspond to the different parameter pairs as follows: Bar strength $A_{2bar}$ and spiral strength $A_{2spiral}$, Bar length $r_{bar}$ and spiral strength $A_{2spiral}$, Bar pattern speed $\Omega$ and spiral strength $A_{2spiral}$, Bar strength $A_{2bar}$ and spiral arm pitch angle $\Psi$, Bar length $r_{bar}$ and spiral arm pitch angle $\Psi$, and Bar pattern speed $\Omega$ and spiral arm pitch angle $\Psi$. }
    \label{TE_total_violin}
\end{figure*}

\begin{figure*}[hbt]
    \begin{center}
    \includegraphics[scale=0.45]{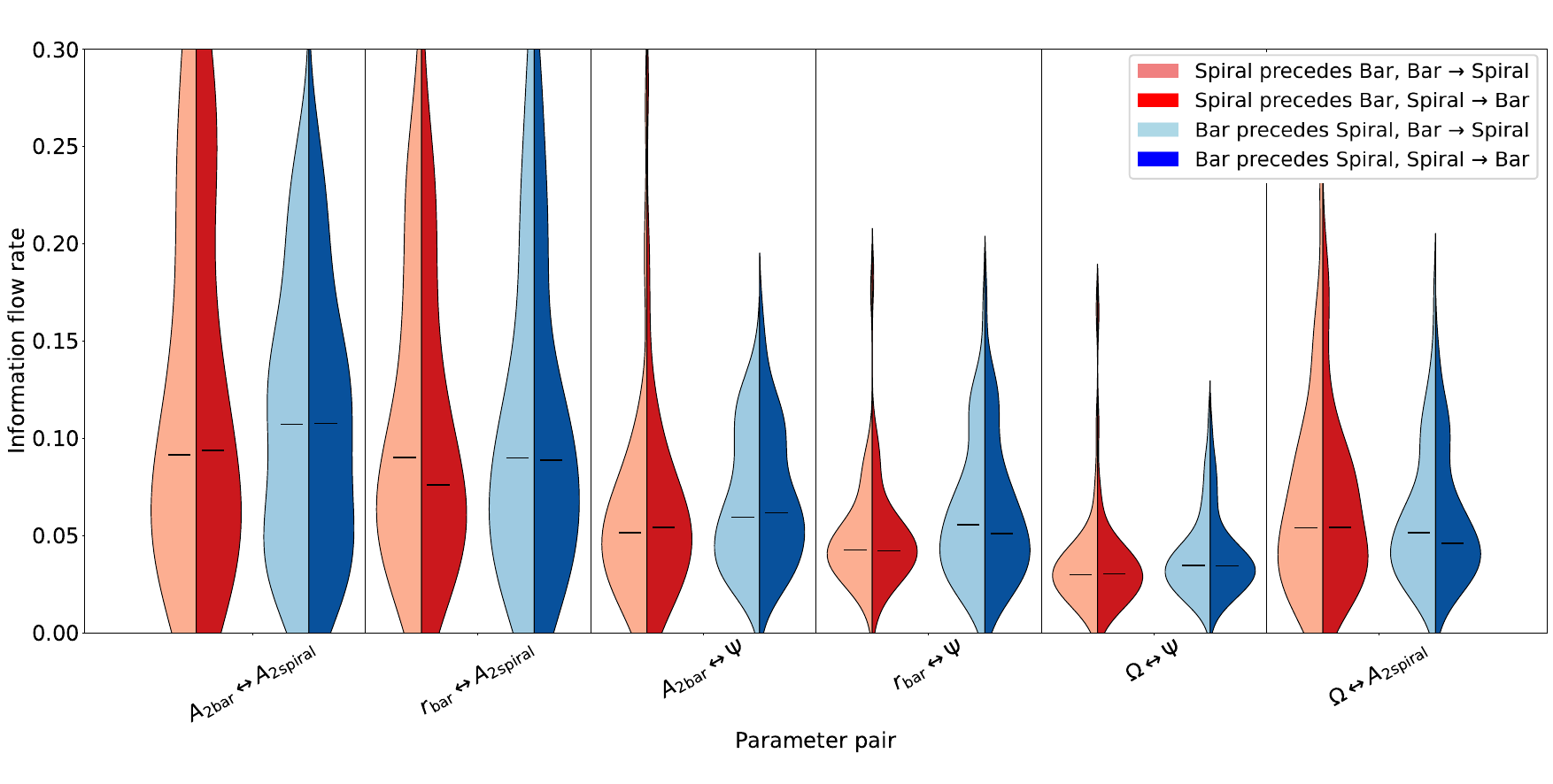}
    \end{center}
\caption{Violin plots of Liang information flow rates for different parameter pairs. The two sides of the violin illustrates the distribution of Liang information flow rate in two opposite directions: Liang information flow rate from spiral to bar ($I_{\text{spiral\ to\ bar}}$) (darker shade) and Liang information flow rate from bar to spiral ($I_{\text{bar\ to\ spiral}}$) (lighter shade). Red shades represent galaxies where spiral precedes bar, while blue shades indicate galaxies where bar precedes spiral. Black horizontal lines inside the violins denote the median value of the Liang information flow rate. Left to right panels correspond to the different parameter pairs as follows: Bar strength $A_{2bar}$ and spiral strength $A_{2spiral}$, Bar length $r_{bar}$ and spiral strength $A_{2spiral}$, and Bar pattern speed $\Omega$ and spiral strength $A_{2spiral}$ and Bar strength $A_{2bar}$ and spiral arm pitch angle $\Psi$, Bar length $r_{bar}$ and spiral arm pitch angle $\Psi$, and Bar pattern speed $\Omega$ and spiral arm pitch angle $\Psi$. Vertical black lines separate the violins for different groups of parameter pairs.}
\label{IF_total}
\end{figure*}

\begin{table*}[hbt]
\centering
\caption{Median values of the Transfer Entropy corresponding to the different pairs of structural and kinematic pairs of bars and their associated spiral arms in our galaxy samples. The errors denote
the differences between the median and the first and third
quantiles.}
\label{Table_TE}

\begin{tabular}{lccc}
\hline\hline
\textbf{Parameter pair} &
\textbf{S}\footnote{Spiral precedes bar} &
\textbf{B}\footnote{Bar precedes spiral} &
\textbf{S$+$B} \\
\hline

\multirow{2}{*}{
$A_{2\mathrm{bar}}$
\shortstack{$\rightarrow$\\$\leftarrow$}
$A_{2\mathrm{spiral}}$
} \footnote{bar strength--spiral arm pitch angle in both directions}
& $0.38^{+0.11}_{-0.11}$ & $0.38^{+0.05}_{-0.06}$ & $0.38^{+0.08}_{-0.06}$ \\
& $0.36^{+0.12}_{-0.10}$ & $0.37^{+0.05}_{-0.04}$ & $0.37^{+0.06}_{-0.05}$ \\

\multirow{2}{*}{
$A_{2\mathrm{bar}}$
\shortstack{$\rightarrow$\\ $\leftarrow$}
$\Psi$
} \footnote{bar strength--spiral arm pitch angle in both directions}
& $0.35^{+0.09}_{-0.09}$ & $0.38^{+0.07}_{-0.09}$ & $0.36^{+0.09}_{-0.08}$ \\
& $0.36^{+0.06}_{-0.07}$ & $0.37^{+0.05}_{-0.07}$ & $0.37^{+0.06}_{-0.07}$ \\

\multirow{2}{*}{
$r_{\mathrm{bar}}$
\shortstack{$\rightarrow$\\ $\leftarrow$}
$A_{2\mathrm{spiral}}$
} \footnote{bar length--spiral amplitude in both directions}
& $0.42^{+0.06}_{-0.07}$ & $0.42^{+0.06}_{-0.08}$ & $0.42^{+0.06}_{-0.08}$ \\
& $0.36^{+0.08}_{-0.08}$ & $0.38^{+0.06}_{-0.08}$ & $0.37^{+0.08}_{-0.09}$ \\

\multirow{2}{*}{
$r_{\mathrm{bar}}$
\shortstack{$\rightarrow$\\ $\leftarrow$}
$\Psi$
} \footnote{bar length--spiral arm pitch angle in both directions}
& $0.42^{+0.05}_{-0.10}$ & $0.38^{+0.10}_{-0.08}$ & $0.39^{+0.08}_{-0.09}$ \\
& $0.38^{+0.07}_{-0.06}$ & $0.35^{+0.08}_{-0.05}$ & $0.36^{+0.07}_{-0.06}$ \\

\multirow{2}{*}{
$\Omega$
\shortstack{$\rightarrow$\\ $\leftarrow$}
$A_{2\mathrm{spiral}}$
} \footnote{bar pattern speed--spiral amplitude in both directions}
& $0.39^{+0.09}_{-0.10}$ & $0.39^{+0.06}_{-0.05}$ & $0.39^{+0.06}_{-0.07}$ \\
& $0.42^{+0.07}_{-0.09}$ & $0.38^{+0.04}_{-0.07}$ & $0.39^{+0.06}_{-0.07}$ \\

\multirow{2}{*}{
$\Omega$
\shortstack{$\rightarrow$\\ $\leftarrow$}
$\Psi$
} \footnote{bar pattern speed--spiral arm pitch angle in both directions}
& $0.38^{+0.06}_{-0.05}$ & $0.35^{+0.06}_{-0.07}$ & $0.36^{+0.06}_{-0.06}$ \\
& $0.37^{+0.04}_{-0.04}$ & $0.33^{+0.09}_{-0.06}$ & $0.35^{+0.06}_{-0.05}$ \\

\hline
\end{tabular}
\end{table*}

\begin{table*}[hbt]
\centering
\caption{Median values of the Liang information flow rate corresponding to the
different pairs of structural and kinematic parameters of bars and their
associated spiral arms in our galaxy samples. The errors denote
the differences between the median and the first and third quantiles.}
\label{Table_IF}

\begin{tabular}{lccc}
\hline\hline
\textbf{Parameter pair} &
\textbf{S}\footnote{Spiral precedes bar} &
\textbf{B}\footnote{Bar precedes spiral} &
\textbf{S$+$B} \\
\hline

\multirow{2}{*}{
$A_{2\mathrm{bar}}$
\shortstack{$\rightarrow$\\ $\leftarrow$}\ 
$A_{2\mathrm{spiral}}$
} \footnote{bar strength--spiral amplitude in both directions}
& $0.09^{+0.11}_{-0.04}$ & $0.11^{+0.04}_{-0.06}$ & $0.10^{+0.07}_{-0.06}$ \\
& $0.09^{+0.12}_{-0.04}$ & $0.11^{+0.04}_{-0.06}$ & $0.10^{+0.08}_{-0.05}$ \\

\multirow{2}{*}{
$r_{\mathrm{bar}}$
\shortstack{$\rightarrow$\\ $\leftarrow$}
$\Psi$
} \footnote{bar strength--spiral arm pitch angle in both directions}
& $0.05^{+0.02}_{-0.02}$ & $0.06^{+0.04}_{-0.02}$ & $0.06^{+0.03}_{-0.02}$ \\
& $0.05^{+0.02}_{-0.02}$ & $0.06^{+0.04}_{-0.02}$ & $0.06^{+0.03}_{-0.02}$ \\

\multirow{2}{*}{
$r_{\mathrm{bar}}$
\shortstack{$\rightarrow$\\ $\leftarrow$}
$A_{2\mathrm{spiral}}$
} \footnote{bar length--spiral amplitude in both directions}
& $0.09^{+0.07}_{-0.04}$ & $0.09^{+0.06}_{-0.04}$ & $0.09^{+0.07}_{-0.04}$ \\
& $0.08^{+0.07}_{-0.03}$ & $0.09^{+0.05}_{-0.04}$ & $0.09^{+0.06}_{-0.04}$ \\

\multirow{2}{*}{
$r_{\mathrm{bar}}$
\shortstack{$\rightarrow$ \\ $\leftarrow$}\
$\Psi$
} \footnote{bar length--spiral aarm pitch angle in both directions}
& $0.04^{+0.01}_{-0.01}$ & $0.06^{+0.02}_{-0.02}$ & $0.05^{+0.02}_{-0.01}$ \\
& $0.04^{+0.01}_{-0.01}$ & $0.05^{+0.03}_{-0.02}$ & $0.05^{+0.02}_{-0.01}$ \\

\multirow{2}{*}{
$\Omega$
\shortstack{$\rightarrow$\\ $\leftarrow$}
$A_{2\mathrm{spiral}}$
} \footnote{bar pattern speed--spiral amplitude in both directions}
& $0.05^{+0.04}_{-0.02}$ & $0.05^{+0.02}_{-0.02}$ & $0.05^{+0.04}_{-0.02}$ \\
& $0.05^{+0.03}_{-0.02}$ & $0.05^{+0.03}_{-0.01}$ & $0.05^{+0.03}_{-0.02}$ \\

\multirow{2}{*}{
$\Omega$
\ \shortstack{$\rightarrow$\\ $\leftarrow$}
$\Psi$
} \footnote{bar pattern speed--spiral arm pitch angle in both directions}
& $0.03^{+0.01}_{-0.01}$ & $0.03^{+0.01}_{-0.01}$ & $0.03^{+0.01}_{-0.01}$ \\
& $0.03^{+0.01}_{-0.01}$ & $0.03^{+0.01}_{-0.01}$ & $0.03^{+0.01}_{-0.01}$ \\

\hline
\end{tabular}
\end{table*}

Using the concept of Transfer Entropy (TE), we next investigate whether any one of the components, namely the bar or the spiral arm, is effectively responsible for driving their co-evolution. To our best possible knowledge, it is the first such attempt in the study of barred-spiral galaxies. We consider $101$ barred
spiral galaxies at a redshift of $z=0$ mentioned in section $\S $ \ref{TNGsamples}.
In Figure \ref{TE_total_violin}, we show the Violin Plots of TE values for different parameter pairs. The two sides of the violin present the distribution of TE in two different directions: TE from spiral-to-bar ($TE_{\text{spiral-\ to-\ bar}}$) (darker shade) and TE from bar-to-spiral ($TE_{\text{bar-\ to-\ spiral}}$) (lighter shade). Red shades represent galaxies where spiral precedes bar, while blue shades indicate galaxies where bar precedes spiral. We calculate the median of the distribution (black lines) to determine the average degree of information transfer from bar (spiral) to spiral (bar), in each case. Each panel represent the TE distributions for a different parameter pair. Table (\ref{Table_TE}) shows the median values of TE for all pairs of parameters for both galaxy samples and the combined sample. \emph{We note, for both the galaxy samples, the median TE value from bar-to-spiral or from spiral-to-bar lies between a common range of $0.33-0.42$, (i) again revealing a strong enough degree of co-evolution between the bar and the spiral arm (ii) equally regulated by both the components}. Further, we also determine the distribution of TE values (from bar-to-spiral as well as spiral-to-bar) by combining the two samples. We found that the median TE values for all parameter pairs are about $0.35-0.42$ in each case, again confirming a fair degree of co-evolution between the bar and the spiral arm, and indicating that both components are equally responsible for driving it. 

Similarly, in Figure \ref{IF_total}, we show Violin plots of IFRs for different parameter pairs. We calculate the median of the distribution (black lines) to determine the average degree of information transfer from bar (spiral) to spiral (bar), in each case. Table (\ref{Table_IF}) shows the median value of Information Flow for all pairs of parameters for both groups of galaxies. We note that for each parameter pair, the median Information Flow values range between $0.03 - 0.11$ for each of the galaxy samples, as well as for the combined samples, confirming the same trend as the TE values {\citep{liang_2015}}.

Finally, we try to assess any subtle yet significant differences in the evolutionary pathways of the two groups of galaxies: those in which bars precede spirals and those in which spirals precede bars. Towards that end, we employ the 2-sample Kolmogorov–Smirnov (KS) test to distinguish the distributions of the TE and the IFR between all pairs of bar and spiral arm parameters. The null hypothesis is that the two samples are drawn from the same distribution. A p-value less than $0.05$ indicates a significant difference between the groups. For TE between all the parameter pairs, only the TE from spiral arm pitch angle to bar pattern speed shows a significant difference between the two groups ($p$ value=$0.02$). For the remaining parameter pairs, the $p$-value is greater than $0.05$, indicating the result is inconclusive. Similarly, the IFR from bar length to spiral arm pitch angle shows a significant difference between the two groups ($p$-value = 0.03). In Figure \ref{AD_TE} (a), we show the cumulative distribution of (i) TE from spiral arm pitch angle to bar pattern speed for two galaxy samples: bar precedes spiral (orange) and spiral precedes bar (blue) (ii) IFR from bar length to spiral arm pitch angle for two galaxy samples: bar precedes spiral (orange) and spiral precedes bar (blue). Interestingly, in the former case, the distribution of the first sample is dominated by low values, and that of the second by high values of TE. In case of the latter, exactly the reverse is noted. This implies that the bar pattern speed is regulated by the spiral pitch angle more strongly when the spiral forms before the bar. Similarly, the bar length regulates the spiral pitch-angle more effectively when the bar forms before the spiral. This complies with the study of \citeauthor{sellwood_1988} (\citeyear{sellwood_1988}) who suggested that within a radius of 1.6 times the length of the bar, spirals act as a continuation of the m=$2$ bar mode but begin to decouple beyond this range. Further, using numerical simulations, \citeauthor{Garma_Oehmichen_2021} (\citeyear{Garma_Oehmichen_2021}) showed that when a bar perturber induces the spirals, the pitch angle best correlates with the bar pattern speed. However, the causal relation between the two parameters remained unexplored.

.

\subsection{Angular Momentum Exchange between the bar and the spiral arm}
The quadrupole moment associated with the gravitational potential of the non-axisymmetric features like the bar and the spiral arms results in torques
. It thereby helps in angular momentum transport (\citeauthor{LyndenBellKalnajs1972} \citeyear{LyndenBellKalnajs1972}, \citeauthor{LyndenBellKalnajs1972} \citeyear{LyndenBellKalnajs1972}). The bar transfers angular momentum to the halo and eventually slows down, which, in turn, strengthens the bar (\citeauthor{Weinberg_1985} \citeyear{Weinberg_1985}, \citeauthor{LyndenBellKalnajs1972} \citeyear{LyndenBellKalnajs1972}, \citeauthor{Athanassoula2005} \citeyear{Athanassoula2005}). The spiral arms transfer angular momentum from one part of the disk to another (\citeauthor{BinneyTremaine1987} \citeyear{BinneyTremaine1987}). \\

To investigate if angular momentum exchange is the primary physical mechanism responsible for the association between the bar and the spiral arm parameters, we first obtain the time series evolution of the angular momentum in the bar and spiral arm region for each one of the 101 samples mentioned in $\S$ \ref{TNGsamples}. We measure TE and IFR between the respective specific angular momentum time series associated with the bar and spiral components. We note a median value of $0.6$ in each direction/case for TE across both galaxy samples. Further, the median IFR values range between 0.2 and 0.3 for both. In fact, \citeauthor{refId0} (\citeyear{refId0}) finds a weak correlation between gravitational torques at the bar end and pitch angle.
Finally, we also employ the Kolmogorov–Smirnov (KS) test to compare the TE of angular momenta in the two galaxy samples. Only the TE from spiral arm angular momentum to bar angular momentum shows a significant difference between the two groups (p value = $0.005$), possibly implying the angular momentum exchange in these two groups of galaxies is different. In Figure \ref{fig:TE_angular}, we show the cumulative distribution of TE from the angular momentum of the bar to the angular momentum of the spiral arm for two galaxy samples: bar precedes spiral (orange) and spiral precedes bar (blue).

 \begin{figure}[hbt]
    \begin{tabular}{c}
        \parbox[c]{0.35\textwidth}{\centering \textbf{(a)}\\
        \includegraphics[width=\linewidth]{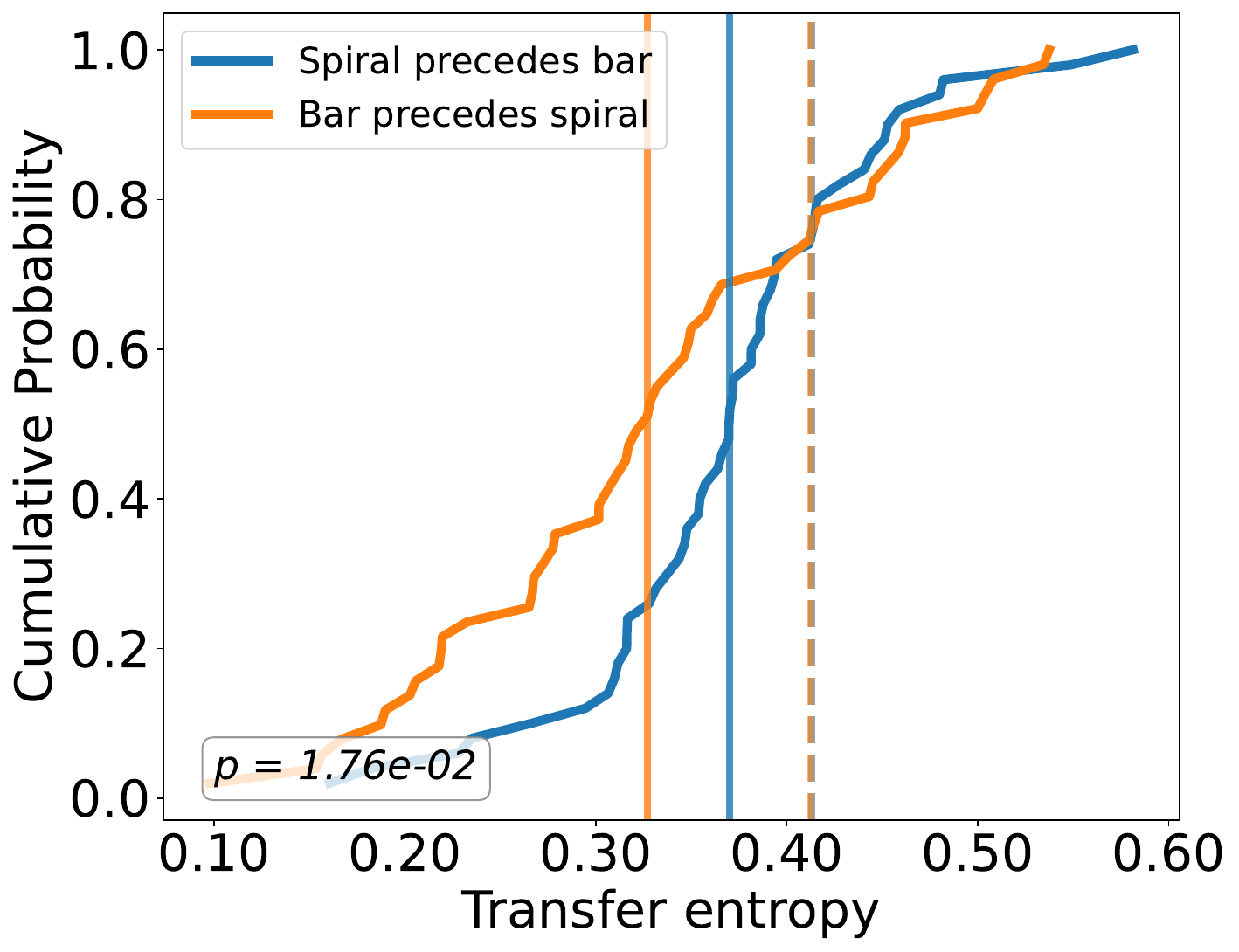}} \\
        \parbox[c]{0.35\textwidth}{\centering \textbf{(b)}\\
        \includegraphics[width=\linewidth]{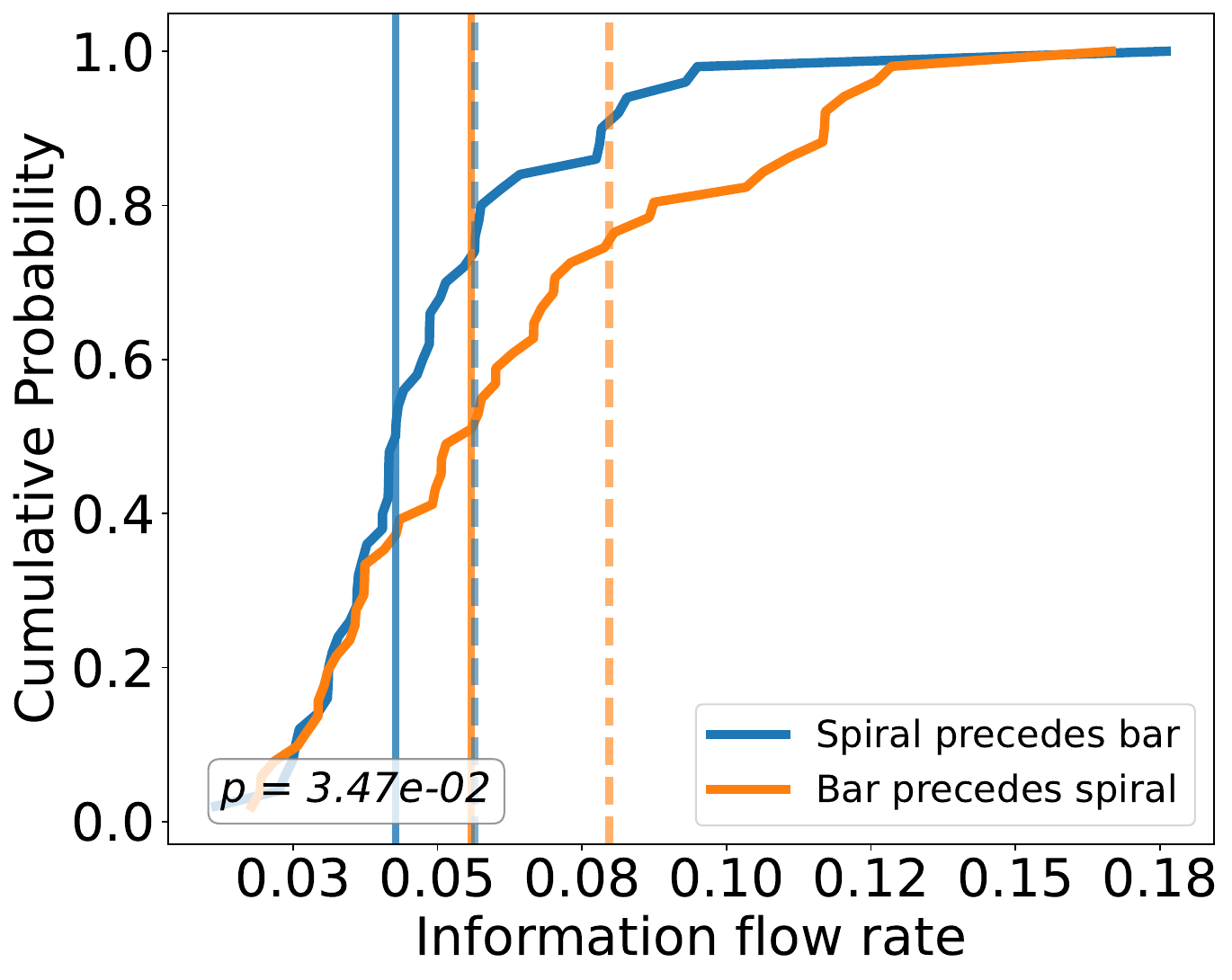}} 
    \end{tabular}
    
\caption{Cumulative distribution of (a) Transfer Entropy from spiral arm pitch angle to bar pattern speed and (b) Liang information flow rate from bar length to spiral arm pitch angle for two galaxy samples: bar precedes spiral (orange) and spiral precedes bar (blue). Solid and dashed lines mark the median and third quartile of each distribution, respectively, and the Kolmogorov–Smirnov test p-value is indicated within the figure.}
\label{AD_TE}
\end{figure}

\begin{figure}
    \centering
    \includegraphics[width=0.75\linewidth]{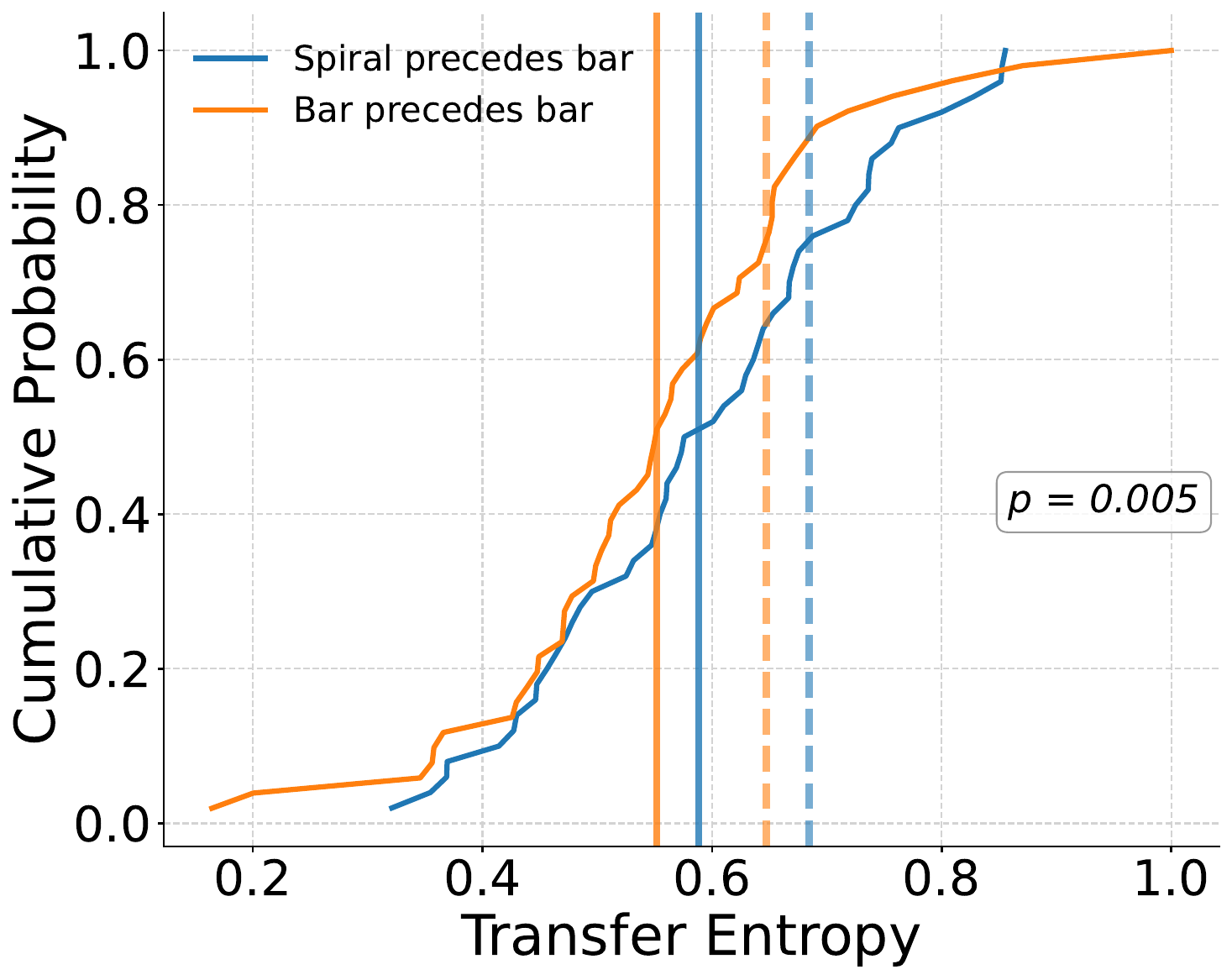}
     \caption{Cumulative distribution of Transfer Entropy from spiral arm specific angular momentum to bar specific angular momentum for two galaxy samples: bar precedes spiral (orange) and spiral precedes bar (blue). Solid and dashed lines mark the median and third quartile of each distribution, respectively, and the Kolmogorov–Smirnov test p-value is indicated within the figure.}
    \label{fig:TE_angular}
\end{figure}

\section{Conclusions}
We investigate the co-evolution of the bar and spiral arms in barred-spiral galaxies using the TNG50 cosmological magneto-hydrodynamical simulation. This analysis focuses on 101 barred-spiral galaxies at a redshift of z = 0, classified into two groups based on whether the bar or the spiral arm appears first in the stellar disk. We note that in the first group, spirals form almost immediately after bars. However, in the second group, bars form about 1.7 Gyrs after the spirals. 
Initially, we calculate the MI between a structural or kinematic parameter of the bar (bar strength $A_{2bar}$, bar length $r_{bar}$, bar pattern speed $\Omega$) and a spiral arm parameter (spiral strength $A_{2spiral}$, spiral arm pitch angle $\Psi$) for the above sample at different $z$ values between $0 < z < 0.5$. Our calculated values of MI for each galaxy sample (0.4-0.6) and the combined sample (0.4-0.8) indicate a significant degree of association between the parameters of the bar and those of the spiral arms. In order to identify if only one of the components is effectively responsible for driving their co-evolution, we next calculate the Transfer Entropy (TE) values (bar-to-spiral TE and spiral-to-bar TE) and the Liang information flow rate (IFR) from the time series data of each of the above bar-spiral parameter pairs. We find that the median bar-to-spiral TE values (0.33 - 0.42) are equal to those of the spiral-to-bar TE values as well as those of the combined sample (0.35-0.42). A similar trend was observed in our calculated IFRs, possibly indicating that the bar and the spiral arm regulate their co-evolution on an equal footing. \\

Further, we employ the Kolmogorov–Smirnov (KS)
to check for subtle yet significant differences between the TE distributions of the two galaxy samples. For the TE between all the parameter pairs, only the TE from bar pattern speed to spiral arm pitch angle shows a significant difference between the two groups (p value = $0.02$). Similarly, the IFR from spiral arm pitch angle to bar length shows a significant difference between the two groups (p-value = $0.03$). Interestingly, in the former case, the distribution of the first sample is dominated by low values, and that of the second by high values of TE. In case of the latter, exactly the reverse is noted. This implies that the bar pattern speed is regulated by the spiral pitch angle more strongly when the spiral forms before the bar. Similarly, the bar length regulates the spiral pitch-angle more effectively when the bar forms before the spiral. \\

Finally, to assess the primary mechanism driving the flow of information, we measure TE and IFR between the angular momenta of the bar and the spiral arm of our sample galaxies. We note a median TE value of $\sim$ 0.6 for both galaxy samples. The median IFR value also ranges between 0.2 - 0.3 for both the galaxy samples, possibly indicating that the flow of information between bar and spiral arm parameters is driven by angular momentum exchange.

\section*{Acknowledgements}
The authors thank Dr Suman Sarkar for sharing details about the Mutual Information study. They also thank Prof. Arif Babul, University of Victoria and the anonymous referee for their valuable comments and suggestions.

\bibliographystyle{mnras}
\bibliography{references} 

\end{document}